\documentclass{IEEEtran}
\pdfoutput=1
\usepackage{amsmath,amssymb,amsfonts}
\usepackage{algorithm}
\usepackage{algorithmicx}
\usepackage{algpseudocode}
\usepackage{graphicx}
\usepackage{textcomp}
\usepackage{subcaption}
\usepackage{multicol}
\usepackage{lipsum}
\usepackage{enumerate}
\usepackage{bm}
\usepackage[numbers]{natbib}
\newtheorem{lemma}{Lemma}
\newtheorem{theorem}{Theorem}
\newtheorem{corollary}{Corollary}

\newtheorem{claim}{Claim}

\newcommand{\RNum}[1]{\uppercase\expandafter{\romannumeral #1\relax}}

\def\BibTeX{{\rm B\kern-.05em{\sc i\kern-.025em b}\kern-.08em
    T\kern-.1667em\lower.7ex\hbox{E}\kern-.125emX}}
\begin{document}

\title{Peaceable Self-Stabilizing Byzantine Pulse Synchronization
\thanks{This work has been submitted to the Springer for possible publication. Copyright may be transferred without notice, after which this version may no longer be accessible.}
}

\author{Shaolin Yu*, Jihong Zhu, Jiali Yang\\Tsinghua University, Beijing, China \\ysl8088@163.com}

\maketitle

\begin{abstract}
For reaching fast and efficient self-stabilizing Byzantine pulse synchronization (SSBPS) upon the bounded-delay message-passing networks, we consider the peaceable SSBPS problem where the resource occupation in the stabilized system is required to be as sparse as possible and the stabilization of the system is required to be as fast as possible.
To solve it, the decoupled absorption process and emergency process are investigated under a general framework.
The constant-time two-stage absorption process and more than one kind of emergency process are provided in integrating the merits of temporal trails, approximate agreements, deterministic and randomized Byzantine agreements, and self-stabilizing protocols.
With this, not only SSBPS but self-stabilizing Byzantine clock synchronization can be fast established and efficiently maintained.
In the optimal-resilient basic solutions, the deterministic linear stabilization time and the stabilized resource occupation are all optimized to those of the underlying primitives, which is optimal in the presence of Byzantine faults under the classical static adversary.
In the hybrid solutions, faster stabilization can be expected with a worst-case deterministic stabilization time.
In all solutions, the stabilized resource occupation is at most at the order of approximate agreement, which is a good property in considering real-world ultra-high-reliable hard-real-time applications where pulse synchronization is expected to be established before all upper-layer functions and to be efficiently maintained when the upper-layer functions are in service.
\end{abstract}


\section{Introduction}
\label{sec:Introduction}
Synchronization is probably the most widely studied system behavior in both natural and man-made systems.
In the realm of distributed computing, synchronization has been intensively investigated from the beginning of the discipline to now.
To see this, the self-stabilization (SS) problem first investigated by \cite{Dijkstra1974} is in the background of \emph{the synchronization task between loosely coupled cyclic sequential processes} \citep{Dijkstra1974}.
And the Byzantine agreement (BA) problem first investigated by \cite{PSL1980} is in the background of \emph{synchronization of clocks, stabilization of input from sensors, and agreement on results of diagnostic tests} \citep{PSL1980} in the SIFT project \citep{Wensley1978SIFT}.
Up to now, the self-stabilizing Byzantine pulse synchronization (SSBPS) problem is still open in several ways \citep{Dolev2014PulseGeneration,Lenzen2019AlmostasEasyas}.
This is not a surprise, though, as SSBPS is at the intersection of three long-pursued virtues in the realm: simple, self-stabilizing, and Byzantine-fault-tolerant.

Firstly, the most striking feature of pulse synchronization (PS) is its very simple characteristics in spite of the complicated \emph{loosely coupled} entities, which brings great easiness for both the overall observation and distributed controls.
But to build man-made systems such as avionics \citep{Hitt2006avionics} and other hard-real-time safety-critical systems (HRT-SCS) \citep{Lala1994Architectural,Kopetz2011Principles} upon such simplicity, the underlying synchronization is also expected to be ultra-high reliable.
For this, today's SSBPS is both featured in the spatial and temporal Byzantine-fault-tolerance (BFT).
In the spatial BFT \citep{Lamport1985Synchronizing}, to maintain the synchronized state of the system, each nonfaulty entity in the system should collect sufficient evidence \citep{halpern1984fault,Dolev1986Approximate,schneider1987understanding} from the remote entities to tolerate a number of permanent Byzantine entities.
And in the temporal BFT, or saying the dense-time SS dimension \citep{Dijkstra1974,Kopetz1992Sparse}, to recover from arbitrary transient disturbance \citep{Kopetz2003Fault,Kopetz2004Hypothesis,Avizienis2004taxonomy}, the nonfaulty entities should be synchronized with arbitrary initial states.

Besides the desired ultra-high reliability, SSBPS is expected to be established in a sufficiently short stabilization time in the background of HRT-SCS.
However, although real-world technologies such as the ever-evolving Ethernet have made great progress in improving communication bandwidth and reducing hardware costs, the maximal message delay experienced in the network can still be significant.
As the SSBPS solutions \citep{DaliotBiological2003,daliot2006self,Daliot2006Linear,DolevPulseBoundedDelay2007,Dolev2014PulseGeneration,Lenzen2019AlmostasEasyas}
often rely on many rounds of handshakes (with echoes, relays, acknowledgments, etc.) between the nonfaulty entities to ensure some specific events being in good order, the practical use of SSBPS is gravely restricted even in the high-bandwidth networks.
As a trade-off, practical ultra-high-reliable communication protocols, such as the ones proposed in the time-triggered architecture \citep{RN448,RN927,RN1345,as6802}, are all built upon some kinds of localized Byzantine filters \citep{RN4551}, such as the central guardians \citep{RN927} or monitor pairs \citep{as6802}, which needs very careful localized design, realization, verification, and validation.

Meanwhile, featured by its very simple characteristics, PS often takes a very special role in real-world distributed applications.
Namely, being the lowest-layer synchronous service, PS is expected to be well established before any other synchronous activities, such as the digital clock synchronization \citep{RN1284,Ben2008Fast,HermanPhase}, TDMA communication \citep{RN448,as6003,Flexray}, high-layer fault-tolerant computing \citep{Kopetz1992Sparse,LenzenCounting2015,LenzenOptimalCounting2015,RN4147,Rybicki2016Near}, etc., being performed.
This special role makes a particular opportunity (and also an urgent obligation) for SSBPS to utilize as many allocable resources as it can in the unsynchronized system.
In real-world applications, as the system should reserve resources not only for PS but also for high-layer applications, there are often plentiful available system resources in establishing the initial pulse synchronization.
However, as the computing entities (often the computers, or saying the nodes) in the SSBPS system are mainly programmed to receive and react to the remote messages, these entities could do nothing but repeat themselves in waiting for the delayed remote messages.
In this situation, the stabilization time is still gravely restricted by the message delays even with ideal network bandwidth.

So, in building practical SSBPS systems, a problem is how to trade the initially unallocated bandwidth (and other resources) for shorter stabilization time and then to maintain the synchronized state of the system with less resource occupation (i.e., the stabilized resource occupation) to make rooms for the deployment of higher-layer applications.

\subsection{Motivation}
In this paper, to make a better balance between the stabilization time and the overall resource occupation of SSBPS, we first aim for deterministic optimal-resilient linear-time stabilization together with sparse stabilized resource occupation, under the classical omniscient but static adversary (would be extended to the so-called restricted dense-time adaptive adversary in later discussion).
In deriving such a deterministic SSBPS (DSSBPS) solution, firstly, the desired PS is expected to be established (and maintained) in the \emph{absorption process} during which all nonfaulty nodes are expected to be synchronized (and stabilized) in tracking some specific \emph{trails}.
And secondly, in case that some nonfaulty nodes cannot participate in such an \emph{absorption process}, some kinds of self-stabilizing deterministic BA (SS-DBA) is built upon prior works to provide an \emph{emergency process}.
To be the preferred and commonly executed primitive, the \emph{absorption process} should be realized with as sparse resource occupation as possible.
And to be the backed-up and one-shot executed primitive, the \emph{emergency process} is allowed to be realized with high resource occupation but must provide a fast system-wide stabilization.
Then, in integrating these two kinds of processes, the \emph{emergency process} should only be executed when necessary.
And once system-wide stabilization is reached, no more \emph{emergency process} would be executed, with which the SSBPS solution is said to be \emph{peaceable}.
As is introduced above, the peaceable DSSBPS (P-DSSBPS) is highly desired in the SSBPS-based HRT-SCS.

Additionally, we would also discuss how faster stabilization can be expected with or without the deterministic worst-case guarantee.
This would be discussed under the same framework of the P-DSSBPS solution, but with some randomized ingredients being added in.
In the literature, although some ingredients of randomized BA (RBA) protocols (such as the one proposed by \cite{FM1997}) can be utilized for reaching expected shorter stabilization time in some randomized SSBPS (RSSBPS) solution \citep{Lenzen2019AlmostasEasyas}, these ingredients are not employed in the \emph{peaceable} way.
Namely, in integrating these ingredients for reaching shorter stabilization time, some high-complexity operations are also involved in the stabilized SSBPS systems, which would generate extra resource occupation (in the message, computation, etc) in the stabilized systems.
There are also RSSBPS solutions (such as the ones proposed by \cite{DolevWelchSelf2004,Dolev2014PulseGeneration}) that do not employ BA-related high-complexity operations.
But it is at the expense of longer randomized stabilization time.
Also, we can see that some randomized ingredients are utilized in some other related problems.
For example, by employing the RBA protocol proposed by \cite{FM1989}, self-stabilizing Byzantine digital clock synchronization is achieved in the synchronous system with expected constant rounds \citep{Ben2008Fast}.
However, the construction of the $K$-clock solution is still recursive, with which a common-coin protocol is involved in every layer of the stabilized beats.
By the way, the open problems posed by \cite{Ben2008Fast} that \emph{if digital clock synchronization solutions can be efficiently transported to the bounded-delay model} \citep{Ben2008Fast} also indicates that the SSBPS problem with the dense-time dimension is quite different from the digital $K$-clock problem with the sparse-time dimension \citep{Kopetz1992Sparse}, let alone the peaceable SSBPS (P-SSBPS) problem.
As far as we know, the integration of the randomized ingredients into the P-SSBPS solutions is still not well understood yet.

\subsection{Main Contribution}
The contribution of this work can be measured in several aspects.
Firstly, in our P-SSBPS solutions, the overall stabilized resource occupation is minimized to that of approximate agreement \citep{Dolev1986Approximate}, which is optimal in the presence of Byzantine faults.
And the stabilized message and computation complexity are even lower than that of the general approximate agreement, since only one $2$-bit message is required to be sent from each node during each \emph{round} (in the semi-synchronous sense) in the stabilized system.
So in this aspect, the overall resource occupation is better than the state-of-the-art deterministic solutions \citep{DolevPulseBoundedDelay2007,Lenzen2019AlmostasEasyas}.

Then, with such optimal stabilized resource occupation, the basic P-DSSBPS solution also reaches optimal-resilient deterministic linear stabilization time with $n>3f$, where $n$ and $f$ are respectively the numbers of all nodes and the allowed Byzantine nodes in the system.
The deterministic stabilization time is optimized in that it just equals the stabilization time of the underlying SS-DBA primitive plus a smaller quantity.
Concretely, denoting the stabilization time of the underlying SS-DBA primitive as $\Delta_\mathtt{stb}$, the overall stabilization time of the P-DSSBPS solution is less than $2\Delta_\mathtt{stb}$, which is also better than that of the state-of-the-art solutions \citep{DolevPulseBoundedDelay2007,Lenzen2019AlmostasEasyas}.

Further, with the added randomized ingredients, the expectation of the stabilization time (or saying the expected stabilization time) of the hybrid P-DSSBPS solution can be further reduced in some non-worst cases.
Meanwhile, the worst-case deterministic stabilization time is still within $2\Delta_\mathtt{stb}$.
This is also better than that of the existing P-SSBPS solutions \cite{Dolev2011PulseGeneration,Dolev2014Rigorously,Dolev2014PulseGeneration} we have known where the expected stabilization time is $O(n)$ with some larger linear-coefficients.
Also, the possible integration of the deterministic \citep{Berman1989optimalconsensus,Toueg1987Fast,Daliot2005protocols,Daliot2006Agreement} and the probabilistic \citep{FM1989,FM1997,Alistarh2018CommunicationEfficient,micali2017,Abraham2019optimal,AbrahamDolev2019} ingredients in solving the P-SSBPS problem provides a new perspective for future exploration.

\subsection{Paper Layout}
In the rest of the paper, the related work is presented in Section~\ref{sec:related}.
The system model and notations are given in Section~\ref{sec:model}, during which the P-SSBPS problem and some basic ideas are also described in words of the \emph{processes} and \emph{trails}.
From Section~\ref{sec:overview} to Section~\ref{sec:emergency}, the basic strategies and realizations of the absorption and emergency processes are presented in a heuristic way.
The main result of the paper is given in Section~\ref{sec:result} and is further extended and discussed in Section~\ref{sec:result2} and Section~\ref{sec:discussion}.
The conclusion and some future directions are given in Section~\ref{sec:conclusion}.

\section{Related Work}
\label{sec:related}
SSBPS is a combination of SS, BFT, and PS.
But it is first worth noting that, perhaps the most common misunderstanding of the SSBPS problem just comes from the original concepts of SS and BFT, since traditionally these two fundamental concepts are all given in the synchronous context where the communication between the nodes are assumed to be synchronized with some underlying timing schemes.
And unfortunately, in traditional HRT-SCS \citep{Lala1994Architectural,Hitt2006avionics}, people often categorize the fault-tolerant schemes into \emph{hardware, software, and timing} \citep{Hitt2006avionics}.
And in providing and validating fault-tolerant timing schemes \citep{Butler1984Validation,Butler1987FaultTolerantCS,RICKY1988avionics}, early solutions often assume that either the timing sources only generate benign faults \citep{Lala1994Architectural,Lala1986A} or some initial synchronization scheme exists \citep{halpern1984fault,LynchWelch1984,Dolev1986Approximate,OptimalToueg1987,Dwork1988Partial,Ramanathan1990Synchronization}.
To be involved in it, later \emph{digital clock} synchronization schemes are often integrated with SS and BFT, but together with the assumption of synchronized underlying pulses \citep{DolevWelchSelf2004,Hoch2006SSDigital,Ben2008SSDigital}.
So here we emphasize that the discussed SSBPS problem is to synchronize the distributed computing nodes in the dense-time world without any aid of initial synchronization nor underlying timing scheme.

To construct ultra-high-reliable HRT-SCS with tolerating both transient system-wide disturbance and permanent malign faults in distributed timing sources, the SSBPS problem is extensively investigated in both theory and practice.
The first optimal-resilient (i.e., with $n>3f$) RSSBPS solution is presented with expected exponential stabilization time by \cite{DolevWelchSelf2004} (the semi-synchronous algorithm) upon bounded-delay peer-to-peer networks.
Almost during the same time, a bio-inspired \citep{Friesen1975Physiological,FriesenSynaptic} optimal-resilient DSSBPS algorithm (named as $\mathtt{BIO}$-$\mathtt{PULSE}$-$\mathtt{SYNCH}$) is proposed by \cite{DaliotBiological2003} upon bounded-delay broadcast networks.
As the required broadcast operation can be simulated in bounded-delay networks without authenticated messages \citep{SrikanthSimulating}, this bio-inspired DSSBPS solution also works upon peer-to-peer networks.
The deterministic stabilization time of $\mathtt{BIO}$-$\mathtt{PULSE}$-$\mathtt{SYNCH}$ is $O(f^3)$, which is better than the expected exponential one provided in \cite{DolevWelchSelf2004}.
Later, in the work of \cite{Daliot2006Linear,DolevPulseBoundedDelay2007}, by employing the SS-DBA protocol (named as $\mathtt{ss}$-$\mathtt{Byz}$-$\mathtt{Agree}$) provided by \cite{Daliot2006Agreement} as a core primitive, optimal-resilient DSSBPS is achieved with linear stabilization time.
However, in providing such a deterministic linear stabilization time, the expense is that the messages and computations needed in establishing and maintaining the synchronization are all at least polynomial to $f$.
The RSSBPS solution provided in \cite{DolevWelchSelf2004}, on the other hand, needs less resource occupation during each \emph{round} in both establishing and maintaining the synchronization, but it is at the expense of a very large number of such \emph{rounds} in reaching stabilization when $n$ is large.
Also, in comparing with the resource occupation of the approximate agreement \citep{Dolev1986Approximate}, this RSSBPS solution still generates some extra message complexity during each \emph{round} in acquiring a sufficient number of clock-value messages from other nodes even in the stabilized states.

Since then, people continuously seek ways to further reduce the stabilization time or the resource occupation or both in tackling the SSBPS problem.
In the work of \cite{Lenzen2019AlmostasEasyas}, a more efficient (in the message, computation, etc) DSSBPS solution is presented by recursively constructing the upper-layer DSSBPS pulsers upon the lower-layer ones with periodically executed deterministic BA (DBA) instances and sparsely distributed resynchronization points.
But as the DSSBPS solutions are all built upon some kind of DBA protocols whose stabilization time and resource occupation are lower-bounded by $f$, it is difficult to further break the linear stabilization time and polynomial complexity barriers in establishing and maintaining the synchronization without introducing some randomness \citep{Rabin1983Randomized,FM1997,King2011Breaking} or some authenticated messages \citep{Dolev1983Authenticated,halpern1984fault,Gupta2010Authenticated,AbrahamDolev2019,Abraham2019optimal}.
In \cite{Lenzen2019AlmostasEasyas}, the first expected-sublinear-time (polylogarithmic) RSSBPS solution is also proposed by periodically executing some ingredient of RBA protocols \citep{FM1997,King2011Breaking}.
But in recursively constructing this RSSBPS solution and the former DSSBPS solution with the sparsely distributed resynchronization points \citep{Lenzen2019AlmostasEasyas}, not only some significant costs of the underlying RBA ingredients are remained in stabilized systems but the overall stabilization time is multiplied with some significant factors.
As far as we know, yet no linear-time DSSBPS nor expected-sublinear-time RSSBPS solution (all state-of-the-art) is provided without periodically executing some kind of BA when the system is stabilized.
This gravely restricts the overall efficiency of the SSBPS systems.

To further reduce the resource occupation, there are easier related solutions, but with either limited BFT dimensions or poor worst-case performance.
For example, traditional clock synchronization \citep{LynchWelch1984} built upon the approximate agreement \citep{Dolev1986Approximate} supports sparse resource occupation, but it does not achieve full temporal BFT and may need a lot of rounds for worst-case convergence even in synchronous systems with unbounded clocks.
And the clock synchronization solutions \citep{OptimalToueg1987,RN4355,5474178} based on simulating the broadcast primitive \citep{SrikanthSimulating} lack the temporal BFT, too.
Other integrated synchronization schemes, such as \cite{OptimalGradientDynamic2010,DOLEV2016929HEX,Gradient2019}, still compromise with limited spatial or temporal BFT.
In \cite{Dolev2014Rigorously,Dolev2014PulseGeneration}, the $FATAL$ protocol can achieve RSSBPS with sparse resources, but it is at the expense of expected-linear-time stabilization.
The $FATAL$ protocol is supposed to be executed in tiny-sized systems running at GHz speed and thus mainly aims for trading the abundant time for the stringent system resources.
In moderate-sized systems, for example, the Ethernet-based distributed embedded systems, practical SSBPS solutions are often built upon low-degree networks with restricted failure-modes \citep{RN927,as6802,Ballesteros2013Towards,YuCOTS2021}.
With our limited knowledge, yet no SSBPS solution makes a good balance in providing both sparse resource occupation and fast stabilization with the full spatial and temporal BFT without taking some restricted failure-mode assumptions.
Especially, when some desired fast stabilization is provided, some form of BA is also periodically executed in the stabilized systems.
And when the expensive BA operations are avoided in the stabilized systems, at best expected-linear-time RSSBPS solutions are provided with the sparsely distributed resynchronization points (which contributes high linear coefficients in the overall stabilization time).

There are also other possible ingredients for constructing better SSBPS solutions.
For example, by combining the secret-sharing-based \citep{ShareSecret1979} eventual RBA \citep{FM1989,FM1997} and various kinds of immediate DBA, the advantages of RBA and DBA can be both acquired in the hybrid BA protocols \citep{Goldreich1990bestbothworlds}).
And with scalable leader elections \citep{King2006}, communication-efficient RBA is also provided with overall sub-quadratic bits both along with polylogarithmic-time with high probability \citep{King2011Breaking,FastBA2013}.
However, as the dense-time PS problem is quite different from the sparse-time \citep{Kopetz1992Sparse} (or saying synchronous) BA problem, the integration of these possible ingredients into the P-SSBPS solutions is nontrivial.
In the next section, we would state the SSBPS problem and the related concepts with the \emph{dense-time} dimension in a formal and self-contained way.
Upon this, all the details in solving the P-SSBPS problem would be heuristically unfolded.
We would find that the so-called \emph{dense time} distinguishes the SSBPS problem from the self-stabilizing Byzantine-fault-tolerant digital clock problem \citep{Ben2008Fast} in that the events generated in the distributed nodes along with the dense time cannot always be \emph{well-separated} \citep{Kopetz2011Principles} in the dense-time processes.

Before proceeding to the details, it is worthy to retrospect the very first RSSBPS solution.
A simple idea behind the semi-synchronous algorithm provided in \cite{DolevWelchSelf2004} is that SSBPS can be achieved by integrating two basic procedures: the \emph{averaging} procedure and the \emph{hopping} procedure \citep{DolevWelchSelf2004}.
The \emph{averaging} procedure just trivially simulates the traditional approximate agreement for maintaining the synchronized state.
The \emph{hopping} procedure, on the other hand, provides critical strategy for \emph{dense-time} SS.
It is shown that the distributed entities can \emph{hop into unison} along with the \emph{dense time} in the presence of the malicious adversary.
However, the integration strategy proposed in \cite{DolevWelchSelf2004} takes the \emph{hopping} procedure as the more frequently executed procedure.
And the realization of the \emph{hopping} procedure is even simpler than the \emph{averaging} procedure.
Under this very intuitive strategy, the complexity of the \emph{hopping} procedure is minimized.
But the \emph{luck} of the never-give-up hopping entities depends on the eventual coincidence of the $n-f$ $s$-faced ($s\geqslant n-2f$) dices tossed in the distributed way, which may takes a very long time to succeed in the presence of the \emph{fair God}.
For reaching faster and more efficient P-SSBPS instead, we take almost the opposite direction.
Roughly speaking, we explore if better SSBPS can be achieved with a more complicated and less frequently performed SS scheme.
With the lovely story of \cite{2015Hoppelpopp}, we call such a peacefully-undesired but emergently-desired SS scheme the \emph{Hoppelpopp} \citep{2015Hoppelpopp}.

\section{System Model and Notations}
\label{sec:model}
\subsection{System, Events and Processes}
Formally, the distributed system $\mathcal{S}$ consists of $n$ nodes, denoted as $N=\{1,2,\dots,n\}$, which are fully connected in a bounded-delay message-passing network $K_n$.
Each node $p\in N$ can generate some instantaneous events and no other event can be generated in $\mathcal{S}$.
Denoting $\mathcal{Y}$ as the set of all event types, each event generated in $\mathcal{S}$ is with a type $y\in \mathcal{Y}$.
For simplicity, we assume that each node $p\in N$ can generate at most one event with the same type at the same instant.
Thus, along with the dense time (defined shortly after), each event generated in $p$ can be uniquely represented as $e(y,t,p)$, where $t\in \mathbb R$ (the real number set) is the instant of generating the event measured in the dense time.
With this, a process $E$ is defined as an event set $e(Y,I,P)=\{e(y,t,p)\mid  y\in Y \land t\in I \land p\in P\}$ with the specific event types $Y\subseteq \mathcal{Y}$, the specific time interval $I\subseteq \mathbb R$ and the specific node-set $P\subseteq N$.
As $E$ is the set of events, $|E|$ is used to denote the number of events in $E$ and we have $E=\emptyset$ iff $|E|=0$.
Particularly, an execution of $\mathcal{S}$ can be represented as the process $E^\chi=e(\mathcal{Y},I^\chi,N)\in \mathbf{E}$, where $I^\chi$ is the whole execution time interval of $E^\chi$ and $\mathbf{E}$ is the set of all possible executions of $\mathcal{S}$.
And for each $p\in N$, $E_p=\{e(y,t,p')\mid e(y,t,p')\in E \land p'=p\}$ is called the local process of $p$ in $E$.
We use $E_P=\cup_{p\in P}E_p$ to denote the process of $P$ in $E$.
Denoting $E[I_1]=\{e(y,t,p)\mid e(y,t,p)\in E \land t\in I_1\}$, we say that $E^\chi_p[I_1]$ is the local process of $p$ during $I_1$ in $E^\chi$.
Also, we use $Nodes(E,I_1)=\{p\mid e(y,t,p)\in E[I_1]\}$ to denote the nodes which generate at least one event during the time-interval $I_1$ in $E$, with which $Nodes(e(Y,I,P),I)$ is shortened as $Nodes(Y,I,P)$.
We say $E$ is generated by $P_1$ during $I_1$ iff $Nodes(E,I_1)\subseteq P_1$.
For $E=e(Y,I,P)$, if $Nodes(Y,I,P)=P$, we say $E$ is globally generated by $P$, simply denoted as $e(Y,I,P)\in \mathbf{E_{gg}}$ when it is not confused.
Similarly, for a node-set $P_1$, we use $Times(E,P_1)=\{t\mid e(y,t,p)\in E \land p\in P_1\}$ to denote the instants at which at least one event is generated by the nodes $P_1$ in $E$, with which $Times(e(Y,I,P),P)$ is shortened as $Times(Y,I,P)$.
For any $Y\subseteq\mathcal{Y}$ and $i\geqslant 1$, we use $Times(Y,I,P)_i$ and $Times(Y,I,P)_{-i}$ to respectively denote the $i$th instant of some nodes in $P$ generating some events $Y$ since and before $I$ (include but not restricted in $I$), with allowing $Times(Y,I,P)_i=+\infty$ and $Times(Y,I,P)_{-i}=-\infty$ if no such event exists.

With this, a node $q$ is nonfaulty during $I$ in $E^\chi$ iff all events in $E^\chi_q[I]$ are generated in the following ways.
Firstly, $q$ should generate the ticking events (typed as $\mathtt{C}$) periodically and count them with a local tick-counter $\tau_q$ (referred to as the local time of $q$).
Denoting $[i:j]=\{i,i+1,\dots,j\}$ as the set of all integers in $[i,j]$, the local time $\tau_q$ should be in some $[0:\tau_{max}-1]$, where $\tau_{max}$ is large enough to run the provided algorithms.
But in considering SS, no matter how large the $\tau_{max}$ is, the local time with the value $\tau_{max}-1$ can overflow and be reset to $0$ at the next $\mathtt{C}$ event.
Also, in considering real-world clocks, the durations between the adjacent $\mathtt{C}$ events (the ticking cycles) in $E^\chi_q$ can fluctuate in a bounded range and can be different in the nonfaulty nodes, if it is measured in the universal reference time.
Here, the universal reference time (i.e. the \emph{dense time}) is assumed to be always at the speed of the slowest local time of the nonfaulty nodes and being without any overflow (like the reference time defined in \cite {Dolev2014PulseGeneration}).
For simplicity but without loss of generality, we take the unit of the reference time $t$ as the largest ticking cycle in the nonfaulty nodes during the discussed execution $E^\chi$ and use $t$ as a real number in $\mathbb R$ (or some subset being dense in $\mathbb R$).
And if not specified, the words \emph{time}, \emph{time interval}, \emph{duration} and \emph{instant} all refer to this reference time.
Upon this, if $e(\mathtt{C},q,t_1)$ and $e(\mathtt{C},q,t_2)$ are two adjacent $\mathtt{C}$ events generated in $q\in Q$, we always have $1 \leqslant 1/|t_2-t_1|\leqslant \vartheta$ with $\vartheta=1+\rho\geqslant 1$ and $\rho\ll 1$ being the maximal relative drift-rate between the nonfaulty nodes.
Generally, to be exact, we have $t_2-t_1\leqslant |e(\mathtt{C},q,[t_1,t_2])|\leqslant \vartheta (t_2-t_1)+1$ for all $0\leqslant t_1\leqslant t_2$ in considering the real-world hardware clocks.
But when $|t_2-t_1|\gg 1$, the difference between $\vartheta |t_2-t_1|+1$ and $\vartheta |t_2-t_1|$ is often ignored for simplicity.
Without loss of generality, here we take this simplification but with the reminder that the additional $1$ tick should be considered in realizing real-world algorithms.

Secondly, $q$ should correctly generate the sending and receiving events (respectively typed as $\mathtt{S}$ and $\mathtt{R}$) in the network $K_n$.
At the same instant $t$, $q$ can send at most one message to each node $p\in N$.
With this, the message being sent from $q$ to $p$ at $t$ can be uniquely represented as $m_{t,q,p}$.
In a nonfaulty bounded-delay network $K_n$, denoting $t_{r}(m)$ as the receiving instant of the message $m$, we have $t_{r}(m_{t_0,q_1,q_2})< t+d_c$ for all $t_0\leqslant t$ and nonfaulty nodes $q_1,q_2$, where $d_c$ is a relaxed communication-delay bound measured in the reference time.
In the uniform solutions where all nonfaulty nodes share the same algorithm, instead of sending the pairwise messages distinctively, $q$ is often required to \emph{distribute} the same value to all the nodes at the same instant.
Namely, denoting the value of the message $m_{t,q,p}$ as $v_{t,q,p}$, when $q$ distributes the value $v_{t,q}=v$ at $t$, we have $v_{t,q,p}=v$ for all $p\in N$.
And as the $\mathtt{R}$ events can be distinguished in each nonfaulty node $p$ with the distinct edges in $K_n$, we assume that $p$ can always identify the senders of the messages.

Thirdly, the local events generated in $q$ should also all be according to the provided algorithms.
Concretely, each algorithm consists of one or several guarded blocks (the top-layer functions) that can be executed as a result of the corresponding guarded events (the top-layer events) being generated.
By denoting an execution of a guarded block $g$ in $q$ as a local process $E_q=e(Y_g,I_g,q)$, the temporally ordered events in $E_q$, which is started with the guarded event $y_g$, should be in accord with the lines of the functions sequentially executed in the block $g$.
Sometimes, to conveniently describe the algorithms, there could be blocking, non-blocking or preemptive executions of the functions and subfunctions (detailed later when needed).
But in all the cases, the overall execution time $|I_g|$ of $E_q$ should always satisfy $|I_g|<d_o$ in the bounded-delay model, where $d_o$ is a relaxed local execution-delay bound measured in the reference time.
Also, $q$ should be in a valid state with which all local constants and local variables defined in the algorithms are in their valid ranges and all possible local inconsistencies of the local variables can be locally corrected in $q$ within a $\Delta_\mathtt{c}$ duration.

On the whole, with the relaxed delay bounds $d_c$ and $d_o$, we can set a relaxed upper bound of the overall message delay (for both the remote messages and the local messages such as the timeouts) as some fixed number $d\geqslant 2d_o+d_c$.
Generally, the distributed controls on the system $\mathcal{S}$ can be viewed as several pairs of causal events, such as the corresponding $\mathtt{S}$ and $\mathtt{R}$ events, the $y_g$ and $y\in Y_g$ events, etc.
In every pair of causal events $e(y_1,t_1,q_1)$, $e(y_2,t_2,q_2)\in E_Q^\chi$ with $t_1<t_2$, $t_2-t_1$ can be viewed as the bounded delay of the causality between the corresponding cause and result events.
With this delayed causality, the reference time is called \emph{dense} \citep{Kopetz1992Sparse} mainly in that for every such pair of causal events, there always exists $t_3\in (t_1,t_2)$ that another event $e(y_3,t_3,q_3)$ can be one of the events in some pair of causal events in the same $E_Q^\chi$.
In other words, different pairs of causal events linked by the \emph{bounded-delay causality} cannot always be well-separated or aligned in the \emph{dense} time.

In considering the BFT problem, by denoting the set of all nonfaulty nodes during $I=[t_0,t_1]$ as $Q$, we say $\mathcal{S}$ is nonfaulty during $[t_0+d,t_1]$ iff $K_n$ is nonfaulty and $|Q|\geqslant n-f$ with $n>3f$ during $I$.
In considering the SS problem, there could be transient disturbances \citep{Kopetz2003Fault,Kopetz2004Hypothesis} during which $Q$ might be $\emptyset$ and the nodes in $F=\{p\mid p\in N\land p\notin Q\}$ can fail arbitrarily in $\mathcal{S}$.
For simplicity and without loss of generality, we assume $\mathcal{S}$ is nonfaulty during $[0,+\infty)$ in $E^\chi$ with each $q\in Q$ being in an arbitrary valid state at the instant $0$.
Then during $[0,+\infty)$, each node $p\in F$ can have an arbitrary (or even undefined) state and be able to make its arbitrarily valued (but still being valid) messages being arbitrarily received (consistently or not) in the nonfaulty nodes at arbitrary instants.
And the nodes in $F$ can ideally collude together under the full control of a malicious adversary who can also arbitrarily choose the instants of all events in $E^\chi$ within the bounded ranges of delays and the relative drift-rates (in $[0,d)$ and $[0,\rho]$, respectively).
In the basic solutions, the adversary is assumed to be omniscient (\emph{non-cryptographic} \citep{BenOr1988NonCryptographic}, \emph{rushing} \citep{King2011Breaking} and having access to even \emph{private channels}) but static.
Other types of adversaries would be given when the related problems are discussed.


Here for convenience, we define the sum and product of two intervals $I$ and $I'$ as respectively $I+I'=\{a+a'\mid a\in I\land a'\in I'\}$ and $II'=\{aa'\mid a\in I\land a'\in I'\}$, where $I$ and $I'$ can take the form of $[a,b]$, $[a,b)$, $(a,b]$ and $(a,b)$ with $a\geqslant 0$.
Also, we do not strictly differentiate between the single-element set $\{a\}$ and $a$.
Further, for any $I$ and $I'$ we use $[I,I']$ (and also the other three forms) as a shorthand of $\{c\mid c\in [a,a'] \land a\in I \cup I' \land a'\in I \cup I' \}$ with $[a,a']=\emptyset$ if $a>a'$.
Besides, as the local times are always in $[0:\tau_{max}-1]$, we denote $\tau_1\oplus \tau_2=(\tau_1+ \tau_2)\bmod \tau_{max}$ and $\tau_1\ominus \tau_2=(\tau_1- \tau_2)\bmod \tau_{max}$ for the local times $\tau_1$ and $\tau_2$.
And for the local time interval $J=[\tau_1:\tau_2]$, we allow not only $\tau_1\leqslant \tau_2$ but $\tau_1>\tau_2$.
In the latter case, the local time interval $J$ can also be represented as $[\tau_1:\tau_{max}-1]\cup[0:\tau_2]$ with the common representation.
Also, the sum of $J$ and $J'$ is defined as $J+J'=\{\tau\oplus \tau'\mid \tau\in J\land \tau'\in J'\}$.

\subsection{Pulsing Process}
The purpose of PS is to synchronize the \emph{pulses} in all nonfaulty nodes of $\mathcal{S}$.
By definition, a \emph{pulse} is a distinguished event (the \emph{pulsing event}, generally typed as $\mathtt{P}$) which can be generated in every node $p\in N$.
In the execution $E^\chi$, each pulse generated in $p\in Q$ can be uniquely represented with a point $(t,p)\in [0,+\infty) \times N$ and can be simply denoted as $e(t,p)=e(\mathtt{P},t,p)$.
Generally, we say a node $p\in N$ is \emph{pulsing} at $t$ in $E^\chi$ iff $e(t,p)\in E^\chi$.
And a pulsing process $E=e(I,P)=\{e(t,p)\mid t\in I \land p\in P\}$ is defined as a subset of $e(\mathtt{P},I^\chi,N)$ in the range of a specific time-interval $I$ and a specific node-set $P$.
With this, a node-set $P_1\subseteq P$ is said to be $(\epsilon,\phi^-,\phi^+)$-synchronized in $E=e(I,P)$, iff
\begin{eqnarray}
\label{eq_sync}
\forall p\in P_1:\forall t\in Times(E,p): \exists I'\subseteq I:t\in I' \land |I'|\leqslant\nonumber\\
 \epsilon \land P_1\subseteq Nodes(E,I') \land p\notin Nodes(E,(t,t+\phi^-)) \land\nonumber\\
 p\in Nodes(E,[t+\phi^-,t+\phi^+])
\end{eqnarray}
holds for some $\phi^->3\epsilon$.
And if $P_1=P$, we say the pulsing process $E$ is $(\epsilon,\phi^-,\phi^+)$-synchronized, denoted as $E\in \mathbf{E}_{\epsilon,\phi^-,\phi^+}^{w}$, where $w=|P|$ is referred to as the width of $E$.
In this case, as the pulsing spans $I'$ in (\ref{eq_sync}) are also well separated, we can define the length of an $(\epsilon,\phi^-,\phi^+)$-synchronized finite pulsing process $E$ as $|Times(E,p)|$ with any $p\in P$.
Thus the $w$-width $l$-length $(\epsilon,\phi^-,\phi^+)$-synchronized finite pulsing process $E$ is denoted as $E\in \mathbf{E}_{\epsilon,\phi^-,\phi^+}^{w,l}$, simplified as $\mathbf{E}_{\epsilon}^{w}$ when $l=1$.

Now, the basic problem of SSBPS (include both the DSSBPS and the RSSBPS) is to provide a desired positive probability in $(0,1]$ of generating $e([t,+\infty),Q)\in \mathbf{E}_{\epsilon_0,T^-,T^+}^{|Q|}$ with $t\leqslant \Delta_\mathtt{stb0}$.
For convenience, the set of all such desired pulsing processes is denoted as $\mathbf{E_0}$.
And generally, by defining a selection function $s:\mathcal{Y}\to \{0,1\}$, a general process $E$ can also be further abstracted as a pulsing process $\bar s(E)=\{e(t,p)\mid e(y,t,p)\in E \land s(y)=1\}$, where $\bar s$ is used as a filter of the typed events.
For example, by defining $s_Y(y)\equiv (y\in Y)$ for every $Y\subseteq \mathcal{Y}$, $\bar s_\mathtt{P}(E^\chi_Q)$ is the pulsing process of $Q$ in the execution $E^\chi$.

\subsection{Trails of Pulsing Process}
As the pulses are all locally generated in the distributed nodes, a nonfaulty node cannot directly observe the pulses generated in other nodes.
To gain an approximate observation of a pulsing process $E=e(I,P)$ with $|P|>1$, each node $p\in Q$ would distribute a \emph{mark} message $m_{t,p}$ with a value $v\in V$ when $e(t,p)\in E$, where $V$ is the value-set of the marks.
In this way, when $p$ generates a pulse $e(t,p)$, each $q\in Q$ can locally observe the \emph{mark} of $e(t,p)$ by receiving the message $m_{t,p,q}$ at an instant $t_{r}(m_{t,p,q})\in[t,t+d)$.
By denoting $\tau_q(t)$ as the local time of $q\in Q$ at instant $t$, $q$ can record this receiving instant as $\tau=\tau_{q}(t_{r}(m_{t,p,q}))$ and record the mark of $e(t,p)$ as a triple $(v,\tau,p)$ in node $q$.
With this, a temporal trail (\emph{trail} for short) can be defined as $\psi_{I,P,q}= \{(v,\tau_{q}(t),p)\mid e(\mathtt{R}_{v,p},t,q)\in e(\mathtt{R},I,q)\land p\in P\}$ with $\mathtt{R}_{v,p}$ being the specific event-type of receiving a mark message with value $v$ from $p$.
Thus, providing that $P\subseteq Q$, every nonfaulty node $q\in Q$ can approximately observe $E=e([t_{1},t_{2}],P)$ as $\{(v_{t,p},\tau_{q}(t_{r}(m_{t,p,q})),p)\mid e(t,p)\in E\}$ in the trail $\psi_{[t_{1},t_{2}+d),P,q}$ before $t_{2}+d$.
Concretely, for any $q\in Q$, denoting $J_{x}^{q}(t)=[\tau_{q}(t)\ominus x :\tau_{q}(t)]$ as the latest $x$-length local time span in $q$ at $t$, we assume that at every instant $t$, $q$ uses only the trails observed within $J_{\Delta}^{q}(t)$, where $\Delta<\tau_{max}$ is a large enough constant.
Denoting the set of all possible trails observed in $J_{\Delta}^{q}(t)$ for all $q\in Q$ as $\Psi$, as the relative drift-rate of $q$ is bounded in $[0,\rho]$, the recorded ticking events in every $\psi\in \Psi$ at $t$ are covered in a time interval $I_{\Delta}(t)=[t-\Delta,t]$.
As $\tau_{max}$ is large enough to differentiate the instants in $I_{\Delta}(t)$, for every instant $t$, $\tau_q(I)=\{\tau_q(t')\mid t'\in I\}$ and $\tau_q^{-1}(J)=\{t' \mid \tau_q(t')\in J\}$ are well defined for all $I\subseteq I_{\Delta}(t)$ and $J\subseteq J_{\Delta}^{q}(t)$.

Similar to the process $E[I]$ during the specific time interval $I$, the subset of trail $\psi$ observed in a specific local time interval $J$ can be denoted as $\psi[J]=\{(v,\tau,p)\mid (v,\tau,p)\in \psi \land \tau\in J \}$.
For convenience, we use $J_{\psi,\delta}$ to denote the last $\delta$ ticks counted in $\psi$ and take $\psi[J_{\delta}]$ as a shorthand of $\psi[J_{\psi,\delta}]$.
Now if $|V|=1$, each mark in the trail $\psi$ can be reduced to $(\tau,p)$.
For $|V|>1$, by taking a specific selection function $b:V\to \{0,1\}$, a multi-valued trail $\psi$ can also be abstracted as a single-valued trail $\bar\psi=\bar b(\psi)\equiv\{(\tau,p)\mid (v,\tau,p)\in \psi \land b(v)=1\}$, where $\bar b$ is a filter of the multi-valued marks.
Generally, we denote $Nodes(\psi,J)=\{p\mid (v,\tau,p)\in \psi [J]\}$ and $Ticks(\psi,P_1)=\{\tau\mid p\in P_1 \land (v,\tau,p)\in \psi\}$ just like that of the general process.
And just like the $(\epsilon,\phi^-,\phi^+)$-synchronized node-set $P_1\subseteq P$ in the pulsing process $E=e(I,P)$, $P_1$ is said to be $(\varepsilon,\varphi^-,\varphi^+)$-aligned in the trail $\psi_{I,P,q}$ iff
\begin{eqnarray}
\label{eq_aligned}
\forall p\in P_1:\forall \tau\in Ticks(\psi,p):\exists J\subseteq \tau_q(I): \tau\in J \land |J|\leqslant \nonumber\\
\varepsilon \land P_1\subseteq Nodes(\psi,J) \land p\notin Nodes(\psi,(\tau,\tau+\varphi^-)) \land\nonumber\\
 p\in Nodes(\psi,[\tau+\varphi^-,\tau+\varphi^+])
\end{eqnarray}
holds for some $\varphi^->3\varepsilon$.
And if $P_1=P\neq\emptyset$, we say $\psi$ is $(\varepsilon,\varphi^-,\varphi^+)$-aligned, denoted as $\psi\in \Psi_{\varepsilon,\varphi^-,\varphi^+}$.
In this case, as $P$ is aligned in $\psi$, we can also define the width of $\psi$ as $|P|$, the length of $\psi$ as $|Ticks(\psi,p)|$ with any $p\in P$, and denote the $i$th ($i\geqslant 1$) latest mark of $p$ in $\psi$ as $\psi^{(i,p)}$.
With this, the set of all $w'$-width $l$-length $(\varepsilon,\varphi^-,\varphi^+)$-aligned trails with $w'\geqslant w$ is denoted as $\Psi_{\varepsilon,\varphi^-,\varphi^+}^{w,l}$, simplified as $\Psi_{\varepsilon}^{w}$ when $l=1$.
For any $\psi\in \Psi_{\varepsilon,\varphi^-,\varphi^+}^{w,l}$ observed in $q$, $\psi$ is respectively called a wide trail and a fellow trail iff $w\geqslant n-f$ and $q\in P$.

In considering the basic SSBPS problem, denoting $\varepsilon_0=\vartheta(\epsilon_0+d)$, $\varphi_0^-=kT^--\vartheta d$ and $\varphi_0^+=\vartheta (kT^++ d)$ with some integer $k\geqslant 1$, a trail $\psi\in \Psi_{\varepsilon_0,\varphi_0^-,\varphi_0^+}$ is called the $k$-stride step trail.
As the adversary can take arbitrary $f$ Byzantine nodes, arbitrary message delays in $[0,d)$ and arbitrary relative drift-rates in $[0,\rho]$, any $q\in Q$ cannot count on observing any trail better than the $(n-f)$-width fellow step trails in any $E^\chi$.

\subsection{The Absorption Process and the Emergency Process}
For P-SSBPS, we investigate two kinds of processes.
The one is the \emph{absorption} process, denoted as $e(\mathtt{A},I,P)$, in which every $q\in Q$ can find some specific trails of a sufficient number of nonfaulty nodes $P\subseteq Q$ in $q$'s latest observation.
With these specific trails, every $q\notin P$ is expected to \emph{join} (or saying to be \emph{absorbed} by) $P$ in a desired duration $\Delta_\mathtt{A}$ for reaching some asymptotic convergence.
The other one is the \emph{emergency} process, denoted as $e(\mathtt{G},I,P)$, in which the nonfaulty nodes would participate in some kind of consensus in a desired duration $\Delta_\mathtt{G}$ for reaching some instantaneous convergence.
Here the event types $\mathtt{A}$ and $\mathtt{G}$ are the general types (or saying the type-sets) of all absorption-related and emergency-related events, respectively.
These two kind of processes are decoupled in that $\mathtt{A}\cap \mathtt{G}=\emptyset$ and only one type of event in $\mathtt{A}$ (or $\mathtt{G}$) can influence the generation of the events in $\mathtt{G}$ (or $\mathtt{A}$).

In considering the resource occupation, as the emergency process is often expensive, the absorption process is preferred.
In each execution $E^\chi\in \mathbf{E}$, we say that $E^\chi_Q$ is \emph{peaceful} in $I$ iff $\bar s_\mathtt{G}(E^\chi_Q[I])=\emptyset$.
And we say the process $E_Q=E^\chi_Q[I]$ is \emph{peacefully synchronized}, denoted as $E_Q\in \mathbf{E_1}$, iff $\bar s_\mathtt{P}(E_Q)\in \mathbf{E_0}$ and $E_Q$ is peaceful in $I$.
With this, the P-SSBPS problem requires that the desired synchronized processes should also be eventually peaceful, i.e., in every $E^\chi\in \mathbf{E}$ there should be a desired probability $\eta\in (0,1]$ (include the deterministic and randomised cases) that $E^\chi_Q[[t,+\infty)]\in\mathbf{E_1}$ holds with some $t\leqslant \Delta_\mathtt{pstb}$, by which we say $\mathcal{S}$ is \emph{peacefully stabilized} (\emph{stabilized} for short in the rest of the paper) at $t$.

To satisfy this \emph{peaceability} property, if $\mathcal{S}$ is stabilized at $t_0$, all nonfaulty nodes should take some measures to prevent the initiation of any new emergency process since $t_0$.
Now as a nonfaulty node cannot count on observing any better trail than a step trail with width $n-f$, if $\psi=\bar b(\psi_{I_{\Delta}(t),P,q})$ is a wide $k$-stride step trail since $t_0$ for some selection function $b$ and some stride $k$, $q$ should try to prevent the generation of a new emergency process.
Concretely, $q$ should not participate in initiating any new consensus since $t_0$.
In this way, if all nodes in $Q$ observe such wide $k$-stride step trails since $t_0$, none of them would participate in initiating any new consensus and thus the consensus is expected to be never successfully initiated since the system is stabilized.
However, as the node $q$ does not know whether a node $p\neq q$ is faulty or not, even if $q$ observes a wide step trail in $\psi_{I_{\Delta}(t),P,q}$ with $|P|=n-f$, the nodes $P$ might contain the faulty nodes.
Also, the bounded message delays and relative drifts in the nonfaulty nodes can always generate new errors when these nodes relay their observations.
Generally, if $q$ observes $\psi\in \Psi_{\varepsilon_i,\varphi_i^-,\varphi_i^+}^{w,l}$ for any $i\geqslant 0$, it might only be the result of a pulsing process in $\mathbf{E}_{\epsilon_{i+1},\phi_{i+1}^-,\phi_{i+1}^+}^{w',l}$ with $w'= w-f$, $\epsilon_{i+1}= \varepsilon_i+d$, $\phi_{i+1}^-= \vartheta^{-1}\varphi_i^--d$ and $\phi_{i+1}^+= \varphi_i^++d$ which might be observed in another nonfaulty node $q'$ as a $w'$-width $(\varepsilon_{i+1},\varphi_{i+1}^-,\varphi_{i+1}^+)$-aligned trail $\psi'$ with $\varepsilon_{i+1}=\vartheta(\epsilon_{i+1}+d)$, $\varphi_{i+1}^-=\phi_{i+1}^--\vartheta d$, $\varphi_{i+1}^+=\vartheta (\phi_{i+1}^++d)$.
In this situation, as the system might have not been stabilized, each node $q'$ is expected to be synchronized with the other nonfaulty nodes by initiating an absorption process with some \emph{narrower} (with width $n-2f$) trail $\psi'$.
And if such absorption can be done in $\Delta_\mathtt{A}$, every node $q'\in Q$ is expected to observe a desired wide $k$-stride step trail within some duration since the completion of the absorption process, with which the system is expected to be stabilized.
In a word, some wide $k$-stride step trail $\psi$ should be the herald of an absorption process, with which $q$ should know that a new emergency process is unnecessary since $q$ observes $\psi$.
And this is the basic setting in designing the absorption process.

\subsection{Main Claim}
\label{subsec:mainclaim}
Formally, we show the existence of a basic optimal-resilient P-DSSBPS solution with the following property.
\begin{claim}
\label{claim_result}
If $\mathcal{S}$ is nonfaulty during $[0,+\infty)$, then $\forall E^\chi\in \mathbf{E}:\exists t\leqslant \Delta_\mathtt{pstb}:E^\chi_Q[[t,+\infty)]\in \mathbf{E_1}$ holds with $\Delta_\mathtt{pstb}=O(f)$.
\end{claim}

From Section~\ref{sec:overview} to Section~\ref{sec:result}, we give such a basic solution and show its correctness.
Besides, we also discuss some extensions of the basic solution in Section~\ref{sec:result2} and Section~\ref{sec:discussion}.

\section{An Overview of the Solutions}
\label{sec:overview}
\subsection{The Bunnies}
We have mentioned that the distributed entities investigated in \cite{DolevWelchSelf2004} can resiliently \emph{hop into unison} in the \emph{dense time} under the \emph{bounded-delay causality}.
Besides its usefulness, the \emph{hopping-in-unison} is by itself a very interesting emergence in the real world, especially with delayed causality in the \emph{dense time}.
Now no matter what these distributed entities are in concrete real-world scenarios, our interest is if and how these entities can \emph{hop} better in exhibiting faster and simpler emergence (or, in the words of SSBPS, less stabilization time and lower stabilized complexity).

For a heuristic discussion, now let us compare these entities, namely, the nonfaulty nodes $Q$, to a group of bunnies \citep{2015Hoppelpopp} who want to exhibit the \emph{hopping-in-unison} behavior to scare some predators away.
For this, each bunny can only make a paw print as its mark at each hopping instant and can observe (with some bounded-delays) the marks of other bunnies whose identities can be told from their locations.
Now by taking the hopping events as the pulses, the desired behavior is assumed as the pulsing processes in $\mathbf{E_0}$.

To prevent the formation of such behavior, there are up-to $f$ badgers (as the node-set $F$) who can well disguise themselves and make arbitrary marks that could be inconsistently observed by the bunnies.
In this situation, as a bunny $q\in Q$ cannot tell if a remote $p\in P$ is a bunny or a badger with the location and marks from $p$, $q$ can only observe the trails of as many \emph{might-be bunnies} as possible to mitigate the worst-case interferences of $F$.
That is, for obtaining as much as possible information of the pulsing process $e(I,Q)$, each $q\in Q$ should observe the trail of $e(I,N)$.
Denoting the whole observation in $q$ at $t$ as $\psi_q(t)$, $q$ can see every $\psi\subseteq\psi_q(t)$ at $t$.

It should be noted that a bunny is not allowed to transform into a badger or vice versa here.
Later in this paper, we would allow these transformations under the dense-time \emph{adaptive} adversary.
Now suppose that a bunny $q$ is always a bunny, to take the advantage of the multi-valued marks, whenever $q$ obtains any useful information from the trail $\psi\in \Psi$, $q$ can put this knowledge in its next paw print which in turn can be observed as new marks in other bunnies.
Certainly, the badgers can also take the advantage of the multi-valued marks by generating arbitrary marks.
And as the errors in estimating the actual process from the local observations of the bunnies would be iteratively accumulated in the relayed observations, it seems that the multi-valued marks have no advantage in reaching faster SSBPS.
Nevertheless, some long-term observations on the temporal trails might help the bunnies in breaking this situation.

\subsection{Hoppelpopp and the Best Bunnies\citep{2015Hoppelpopp}}
\label{subsec:Hoppelpoppand}
To utilize the temporal trails, as the first idea, if a bunny $q$ sees some \emph{specific} (here let us view it as a black-box) trail $\psi$ in its latest observation, $q$ would claim itself as the \emph{best} bunny by tagging the marks of its following pulses with $\mathtt{BEST}$.
In this case, we say such a specific trail $\psi$ belongs to the shortcut trails, denoted as $\Psi_\mathtt{short}$.
And a bunny $q$ is a (self-claimed) best bunny iff $q$ sees a trail $\psi\in\Psi_\mathtt{short}$.
As the bunnies cannot reach a fast agreement on a single bunny being the best one, there might be more than one best bunny at the same time.
And the badgers can also claim themselves as the best bunnies too.

Then, for the peaceability property, when a bunny $q$ sees some other \emph{specific} trail $\psi$ made by a sufficient number of best bunnies in its latest observation, $q$ should do nothing but expect the system to be synchronized with an absorption process.
In this case, we say such a specific trail belongs to the happy trails, denoted as $\Psi_\mathtt{happy}$.
And a bunny $q$ is \emph{happy} iff $q$ sees a trail $\psi\in \Psi_\mathtt{happy}$.
For the absorption process, when a bunny $q$ sees some corresponding trail $\psi$ made by a corresponding number of best bunnies in its latest observation, $q$ should be \emph{engaged} in running a new absorption process.
In this case, we say such a corresponding trail belongs to the engaging trails, denoted as $\Psi_\mathtt{engag}$.
And a bunny $q$ is \emph{engaging} iff $q$ sees a new trail $\psi\in \Psi_\mathtt{engag}$.
Also, for each $q\in Q$, to avoid the best bunnies making significantly inconsistent observations on the trail of $q$, $q$ would claim itself as a \emph{good} bunny by tagging the marks of its following pulses with $\mathtt{GOOD}$ iff $q$ thinks it is with sufficient accuracy, i.e., the local time passed in $q$ between its current pulse and the previous pulse is in some desired range.
In this case, we say such a trail of $q$ in $q$ belongs to the good trails of $q$, denoted as $\Psi_\mathtt{good}$.
And a bunny $q$ is \emph{good} iff $q$ sees a trail $\psi\in \Psi_\mathtt{good}$.
With this, the shortcut trails should be observed from the trails of the \emph{good} bunnies.
With these, if $q$ is happy, $q$ should know that all bunnies would be engaged in running an absorption process.
In this situation, the remaining problem is how to design the desired absorption process with the good, the engaged, and the best bunnies and the \emph{specific} trails.

Otherwise, if $q$ is not happy, $q$ is not sure whether all bunnies would be engaged and then to be absorbed or not.
In this situation, $q$ expects \emph{Hoppelpopp} \citep{2015Hoppelpopp} the giant grey rabbit (to be different from the white ones \citep{Moreira2009White}) coming to help the bunnies within call.
Here we do not care what the \emph{Hoppelpopp} is and how it comes but only know that the call for the \emph{Hoppelpopp} would generate some unpeaceful events which bring the bunnies pain.
But the gain is that, if the \emph{Hoppelpopp} comes with good luck, it would make a public appearance for all bunnies, with which all bunnies can reach the desired unison by adjusting their paces according to the consistent appearance events (typed as $\mathtt{H}$, $\mathtt{H}\in \mathtt{G}$) globally generated in a duration no longer than $\epsilon_\mathtt{H}$.
But if \emph{Hoppelpopp} comes with its bad luck, only some (at least one but not all) of the bunnies would perceive the $\mathtt{H}$ events, with which the bunnies can still be misled by the inconsistent $\mathtt{H}$ events.
So in the name of peace, the appearance of \emph{Hoppelpopp} is unanimously undesired when all bunnies are happy.
But when all bunnies are unhappy, the appearance of \emph{Hoppelpopp} is unanimously desired.

\subsection{The Strategy for Survival}
\label{subsec:basic-Strategy}
Following this simple idea, each bunny $q$ runs the $\mathtt{Bunny}$ algorithm shown in Fig.~\ref{algo:bunny-basic}.
Here, we use \emph{distribute} to represent the message-distributing primitive without reliable broadcast channels.
Also, we logically use \emph{call} and \emph{wait} to represent the non-blocking executing and the blocking waiting primitives, respectively.
Namely, the \emph{call} primitive can return as fast as possible (before the return of the called function).
While the \emph{wait} primitive would not return before the waited ticks being passed, just like the common functions that are executed without the \emph{call} primitive.
Being viewed as black boxes, the functions \emph{Mark}, \emph{Absorb}, \emph{Engage}, \emph{Help}, etc. should be further realized in concrete solutions, with which the definitions of the trail-sets $\Psi_\mathtt{good}$, $\Psi_\mathtt{short}$, $\Psi_\mathtt{engag}$, $\Psi_\mathtt{happy}$ and the way to generate the events $\mathtt{P}$ and $\mathtt{H}$ should also be given accordingly.
For simplicity and without loss of generality, we assume $q$ can observe a new trail when a new $\mathtt{R}$ or $\mathtt{C}$ event is generated in $q$.
It should be noted that in general, there are many alternatives in realizing all the black-box functions.
Among all of them, the realization given here aims to be as simple as possible to exhibit the core principles.

\alglanguage{pseudocode}
\algrenewcommand{\algorithmiccomment}[1]{\hskip1em//#1}
\begin{figure}[htbp]
\centering
\begin{subfigure}{.49\textwidth}
\begin{algorithmic}[1]
\Statex \underline{when $q$ pulses at local time $\tau_q$}:\Comment{$\mathtt{P}$}
    \State isGood:=($\psi_q(t)\in\Psi_\mathtt{good}$);
    \State distribute \emph{Mark}~($\tau_q$);\Comment{$\mathtt{S}$} \label{code_line_bunny_distribute}
    \State call \emph{Absorb}~($\tau_q$);\Comment{makes an absorption}
\Statex \underline{when $q$ observes a new trail $\psi_q(t)$}:\Comment{$\mathtt{R}$ and $\mathtt{L}_i$}
    \State isBest:=($\psi_q(t)\in\Psi_\mathtt{short}$);
    ~ isHappy:=($\psi_q(t)\in\Psi_\mathtt{happy}$);
    \If{exists new $\psi_1\subseteq \psi_q(t):\psi_1\in\Psi_\mathtt{engag}$}
        \State call \emph{Engage};\Comment{makes an engagement}
    \EndIf
    \If{not isHappy}
         ~call \emph{Help};
    \EndIf
\Statex \underline{when $\mathtt{H}$ happens at local time $\tau_q$}: \label{code_line_hop_appear}
    \Statex ~\Comment{makes an appearance of \emph{Hoppelpopp}:}
    \State \emph{AdjustByHoppelpopp}~($\tau_q$);\Comment{$\mathtt{D_2}$} \label{code_line_hop_adj}
    \State cancel all pending functions;
    \State clear all marks and not record any mark for $\delta_3$ ticks; \label{code_line_hop_spare}
\end{algorithmic}
\caption{The $\mathtt{Bunny}$ algorithm for each $q\in Q$}
\label{algo:bunny-basic}
\end{subfigure}
\hfill
\begin{subfigure}{.49\textwidth}
\begin{algorithmic}[1]
\Statex \underline{\emph{Mark}~($\tau_q$)}:
    \State $v:=\emptyset$;
    \If{isGood}
        $v:=v\cup\{\mathtt{GOOD}\}$;
    \EndIf
    \If{isBest and $k_\mathtt{A}=0$}
        $v:=v\cup\{\mathtt{BEST}\}$;
    \EndIf
    \State \Return{$v$};

\Statex \underline{\emph{Absorb}~($\tau_q$)}:\Comment{can be preemptively canceled}
    \If{$k_\mathtt{A}>0$}
        \State wait $\delta_0$ ticks;
        \State \emph{AdjustByAllFellows}~($\bar b_0(\psi_q(t)[J_{2\delta_0+2\vartheta d}])$);\Comment{$\mathtt{D_0}$}  \label{code_line_absorb_adjust}
        \State $k_\mathtt{A}:=(k_\mathtt{A}+1)\bmod K_\mathtt{A}$;  \label{code_line_absorb_k}
    \EndIf
\Statex \underline{\emph{Engage}}:\Comment{can preemptively cancel \emph{Absorb}}
    \State wait $\delta_1$ ticks;
    \State \emph{AdjustByBestBunnies}~($\bar b_2(\psi_q(t)[J_{\delta_1+\varepsilon_1+2\vartheta d}])$);\Comment{$\mathtt{D_1}$}  \label{code_line_engage_adjust}
    \State wait $\delta_2$ ticks; ~$k_\mathtt{A}:=1$;  \label{code_line_engage_k}
\end{algorithmic}
\caption{A partial realization of the absorption process}
\label{algo:bunny-partial}
\end{subfigure}
\caption{The basic strategy and a partial realization.}
\label{fig:algo_bunny}
\end{figure}

In Fig.~\ref{algo:bunny-partial}, the functions \emph{Mark}, \emph{Absorb} and \emph{Engage} are partially realized for the absorption process.
Firstly, it should be noted that in the function \emph{Absorb}, only the engaged $q$ (i.e., with the absorption counter $k_\mathtt{A}>0$) is absorbed by executing the function \emph{AdjustByAllFellows} after $\delta_0$ ticks waited in executing \emph{Absorb}.
While the engaging $q$ (i.e., with $k_\mathtt{A}=0$) would not be affected by the \emph{Absorb} function.
The reason is that the engagements would be handled exclusively by calling the \emph{Engage} function with a higher preemptive priority.
Namely, whenever $q$ calls the \emph{Engage} function, all possible executions of \emph{Absorb} (and also any previously called \emph{Engage}) would be preempted and canceled by \emph{Engage}.
And even when new \emph{Absorb} is called during \emph{Engage} being executed, this new \emph{Absorb} would still be canceled before it can ever be executed.
Actually, when the rule of preemptive priority is well-realized, the special case of $k_\mathtt{A}=0$ needs not to be explicitly excluded in the \emph{Absorb} function.
It is shown here just for emphasis and clarity.
On the whole, this preemptive rule also applies to the appearance of \emph{Hoppelpopp} which is with the highest preemptive priority.
And during the execution of the response function of the $\mathtt{H}$ event, all pending functions would be canceled and all recorded trails would be cleared.
It should be noted here that when we say some function $x$ is preempted by some function $y$, we mean that the currently executing $x$ and all currently pending $x$ are all preemptively canceled, i.e., all preemptions used in this paper are also with cancellations.

Except for these special cases, the executions of the common functions are assumed to be blocking and non-preemptive, i.e., a common function would not return before all its lines (segmented by the semicolons) being sequentially executed.
And if there is a function being executed (not in waiting) while some new top-layer event (include the events underlined in Fig.~\ref{algo:bunny-basic} and the \emph{call} event) is generated, the new response function would be queued as a pending function before the currently executed function returns.
And if there is more than one pending function at the same time, their execution order can be arbitrarily scheduled as long as the overall message delays are all bounded within $d$.
Among the common functions, it should be noted that in the function \emph{Mark}, an engaged $q$ would not tag its mark $v$ with $\mathtt{BEST}$ even if $isBest$ is true in $q$.
This is also for the special handling of the engagements, as is detailed later.
And on the whole, there are three kinds of time adjustments: the absorptions, the engagements, and the appearances, which are respectively implemented by executing the common functions \emph{AdjustByAllFellows}, \emph{AdjustByBestBunnies} and \emph{AdjustByHoppelpopp}.
In these three adjustment functions, the pace of $q$ is adjusted by resetting the scheduling timer $\tau_{sch}$ whose value indicates the next local pulsing instant of $q$.
And for all $v\in V$, we define $b_0(v)\equiv 1$, $b_1(v)\equiv(\mathtt{GOOD}\in v)$ and $b_2(v)\equiv(\{\mathtt{GOOD},\mathtt{BEST}\}\subseteq v)$.
With this, the corresponding event types of distributing (with line \ref{code_line_bunny_distribute} of Fig.~\ref{algo:bunny-basic}) and receiving the value $v$ from $p$ with $b_i(v)=1$ are respectively denoted as $\mathtt{S}_i$ and $\mathtt{R}_{i,p}$.
The events of observing a new engaging trail and a new happy trail (i.e., with $\psi\subseteq \psi_q(t)[J_{\varepsilon_0}]$ satisfying $\bar b_2(\psi)\in \Psi_{\varepsilon_0}^{n-f}$) are respectively typed as $\mathtt{L_1}$ and $\mathtt{L_2}$.
The $\mathtt{P}$ event generated with $k_\mathtt{A}=k$ (here we assume $K_\mathtt{A}\geqslant2$) is specifically typed as $\mathtt{P}_k$, where $\mathtt{P}=\bigcup_{k\in[0:K_\mathtt{A}-1]}\mathtt{P}_k$ is the generalized type of pulses.
The event of setting $k_\mathtt{A}=k$ is typed as $\mathtt{K}_k$.

Now, for this basic strategy to be effective, two things should be ascertained: how the bunnies are absorbed and what is the \emph{Hoppelpopp}?
These are respectively the tasks of the absorption process and the emergency process, as is detailed in the next two sections.

\section{The Absorption Process}
\label{sec:absorpion}
We first discuss the absorption process.
For clarity, we assume $e(\mathtt{H},I,Q)=\emptyset$ for all $I$ referred in this section.
Later in this paper, we will show how to integrate the absorption processes with the appearance of \emph{Hoppelpopp}.

\subsection{Realization of the Absorption Process}
\label{subsec:realize-Absorption}
A concrete realization of the absorption process is shown in Fig.~\ref{fig:algo_absorption}.
Firstly, in Fig.~\ref{algo:bunny-trails}, with the defined $b_i$ for $i\in\{0,1,2\}$, $q$ thinks it is good when the local time passed between the current and the previous pulses of $q$ is in $[T-\varrho_1,T+\varrho_1]$, where $\varrho_1$ is called the local accuracy threshold.
And $q$ thinks it is among the best ones when $q$ observes $n-f$ marks (include the mark of $q$) with the $\mathtt{GOOD}$ tag in a sufficiently short local time span.
Also in Fig.~\ref{algo:bunny-trails}, $q$ would be happy and be engaged iff $q$ observes respectively $n-f$ marks and $n-2f$ marks with the $\{\mathtt{GOOD},\mathtt{BEST}\}$ tag in sufficiently short local time spans, where $\vartheta\Delta_\mathtt{A}\geqslant \varphi_0^+$ with $\varphi_0^+=\vartheta (K_\mathtt{A} T^++ d)$ and $\varphi_0^-=K_\mathtt{A} T^--\vartheta d$.
In Fig.~\ref{algo:bunny-adjustments}, denoting the event types of the time-adjustments (generated at the completion instant of rescheduling $\tau_{sch}$) in executing the \emph{AdjustByAllFellows}, \emph{AdjustByBestBunnies} and \emph{AdjustByHoppelpopp} functions as $\mathtt{D_0}$, $\mathtt{D_1}$ and $\mathtt{D_2}$, we say $q$ makes an absorption, an engagement and an appearance at $t$ iff $e(y,t,q)\neq \emptyset$ for $y=\mathtt{D_0}$, $y=\mathtt{D_1}$ and $y=\mathtt{D_2}$, respectively.
Notice that we have $|Ticks(\psi,q)|=1$ in executing the \emph{AdjustByAllFellows} function in $q\in Q$.
And the \emph{FTAverage} function can always be correctly executed by setting $\delta_\epsilon=2\delta_0+3\vartheta d$ in each $q\in Q$ in Fig.~\ref{algo:bunny-fta}.
For analyzing the algorithms, we use $x^{(q)}(t)$ to represent the value of the local variable $x$ of $q\in Q$ at the instant $t$, where $x$ can be $isGood$, $isBest$ and so on.

\alglanguage{pseudocode}
\algrenewcommand{\algorithmiccomment}[1]{\hskip1em//#1}
\begin{figure}[htbp]
\begin{subfigure}{.49\textwidth}
\begin{eqnarray}
\Psi_\mathtt{good}=\nonumber\\
\{\psi\in\Psi\mid T-\varrho_1\leqslant \bar b_0(\psi)^{(1,q)}\ominus \bar b_0(\psi)^{(2,q)}\leqslant T+\varrho_1\}\\
\Psi_\mathtt{short}=\nonumber\\
\{\psi\in\Psi\mid \exists \psi_1\subseteq \psi[J_{\varepsilon_0+T+\varrho_1}] :\bar b_1(\psi_1)\in \Psi_{\varepsilon_0}^{n-f} \land \nonumber\\
Ticks(\bar b_1(\psi_1)[J_{\varepsilon_0}],q)\neq \emptyset\}\\
\Psi_\mathtt{happy}=\nonumber\\
\{\psi\in\Psi\mid \exists \psi_1\subseteq \psi[J_{2\varepsilon_0+\vartheta\Delta_\mathtt{A}}] :\bar b_2(\psi_1)\in \Psi_{\varepsilon_0}^{n-f} \}\\
\Psi_\mathtt{engag}=\{\psi\in\Psi\mid \bar b_2(\psi)\in \Psi_{\varepsilon_1}^{n-2f} \}
\end{eqnarray}
\caption{The four kinds of specific trails}
\label{algo:bunny-trails}
\end{subfigure}
\begin{subfigure}{.49\textwidth}
\begin{algorithmic}[1]
\Statex \underline{\emph{AdjustByAllFellows}~($\psi$)}:\Comment{$\mathtt{D_0}$}
    \State $\tau_{sch}:=\tau_{sch}\oplus \emph{FTAverage}(\psi)\ominus Ticks(\psi,q)\oplus T$;
\Statex \underline{\emph{AdjustByBestBunnies}~($\psi$)}:\Comment{$\mathtt{D_1}$}
    \State $\tau_{sch}:=\emph{FTAverage}(\psi)\oplus T$;
\Statex \underline{\emph{AdjustByHoppelpopp}~($\tau_q$)}:\Comment{$\mathtt{D_2}$}
    \State $\tau_{sch}:=\tau_q\oplus T$;
    ~ $k_\mathtt{A}:=1$;
\end{algorithmic}
\caption{The three kinds of adjustments}
\label{algo:bunny-adjustments}
\end{subfigure}
\begin{subfigure}{.49\textwidth}
\begin{algorithmic}[1]
\Statex \underline{\emph{FTAverage}~($\psi$)}:
    \State $\tau_0:=\tau_q(t)\ominus \delta_\epsilon$;
    \State $S:=\emptyset$;\Comment{$S$ is an ascending multi-set of ticks}
    \ForAll {$(\tau,p)\in\psi \land |Ticks(\psi,p)|=1$}
    \State input $S$ with $\tau\ominus \tau_0$;
    \EndFor \Comment{we have $|S|>f$ here}
    \State \Return $\tau_0\oplus ((S^{(f+1)}+ S^{(\min\{|S|,n-f\})})/2)$;
\end{algorithmic}
\caption{The general FTA function}
\label{algo:bunny-fta}
\end{subfigure}
\caption{A realization of the absorption process.}
\label{fig:algo_absorption}
\end{figure}

\subsection{Analysis}
Firstly, the readers might wonder why an engaged bunny would not claim itself to be in the best ones.
Also, why the bunnies should make observations on some $k$-stride trails with $k>1$ rather than just the simple $1$-stride trails?
Here the story can be told from a happy bunny $q$.
Now suppose that $q$ sees a new happy trail at $t$.
Then $q$ knows that all bunnies (also include $q$) at least make their engagements in a coarsely coordinated way, as is shown in the following lemma.

\begin{lemma}
\label{lemma_steadfast}
If $e(\mathtt{L_2},t,Q)\neq \emptyset$, then both $e(\mathtt{D_1},I,Q)\in\mathbf{E_{gg}}$ and $e(\mathtt{L_1},I',Q)=\emptyset$ hold for $I=[t-\epsilon_1+\vartheta^{-1}\delta_1,t_1+\delta_1+2d)$, $I'=[t_1,t+\vartheta^{-1}T-\varrho_1-2\epsilon_1)$ and $t_1=t+\epsilon_1+\epsilon_2+d$.
\end{lemma}
\begin{IEEEproof}
As $e(\mathtt{L_2},t,Q)\neq \emptyset$, by definition there is $q\in Q$, $\psi_1\subseteq \psi_q(t)[J_{\varepsilon_0}]$ satisfying $\bar b_2(\psi_1)\in \Psi_{\varepsilon_0}^{n-f}$ and thus $e(\mathtt{S_2},I_{\epsilon_1}(t),P)\in \mathbf{E}_{\epsilon_1}^{n-2f}$ holds for some $P\subseteq Q$.
Thus, for every $p\in Q$ there exists $\psi_2\subseteq \psi_p(t_2)[J_{\varepsilon_1}]$ satisfying $\bar b_2(\psi_2)\in \Psi_{\varepsilon_1}^{n-2f}$ for some $t_2\in[t-\epsilon_1,t+d)$.
As $\psi_2\in \Psi_\mathtt{engag}$, $e(\mathtt{D_1},I,p)\neq \emptyset$ would hold if $e(\mathtt{L_1},[t_1,t_1+\delta_1+2d),p)=\emptyset$.
Now suppose there exists $t_2'\in I'$ and $\psi_2'\subseteq\psi_p(t_2')[J_{\varepsilon_1}]$ satisfying $\psi_2'\in \Psi_\mathtt{engag}$.
Thus, $t_1<t+\vartheta^{-1}T-\varrho_1-2\epsilon_1$ holds and by definition $\psi_2'$ contains the $\mathtt{BEST}$ and $\mathtt{GOOD}$ marks of at least $f+1$ distinct nodes.
So there exists $p_1\in Q$ generating $e(\mathtt{S_2},t_3,p_1)$ for some $t_3\in[t_2'-\epsilon_2,t_2']$ with $isGood^{(p_1)}(t_3)=isBest^{(p_1)}(t_3)=1$.
So some $\psi_3\subseteq \psi_{p_1}(t_3')[J_{\varepsilon_0}]$ satisfies $|Nodes(\bar b_1(\psi_3),J_{\psi_3,\varepsilon_0})|\geqslant n-f $ for $t_3'\in t_3+(-d,0]$ or $t_3'\in t_3+T[-1,-\vartheta^{-1}]+\varrho_1[-1,1]+[0,\epsilon_1]$.
Thus, there exists $P'\subseteq Q$ satisfying $|Nodes(\mathtt{S_1},I_0',P')|=|P'|\geqslant n-2f$ for $I_0'=I_1'\cup I_2'$ with $I_1'=t_3+(-\epsilon_1-d,0]=(t_1-\epsilon_1-\epsilon_2-d,I']$ and $I_2'=I_1'+T[-1,-\vartheta^{-1}]+\varrho_1[-1,1]+[0,\epsilon_1]$.
With $e(\mathtt{S_2},I_{\epsilon_1}(t),P)\in \mathbf{E}_{\epsilon_1}^{n-2f}$, we have $|Nodes(\mathtt{S_1},I_0,P)|\geqslant n-2f$ for $I_0=[t-\epsilon_1,t]$.
Noticing that $I_0\cap I_0'=\emptyset$ and a good bunny cannot pulse both in $I_0$ and $I_0'$, we have $P\cap P'=\emptyset$.
Thus, denoting $I_0''=I_0\cup I_0'$, we have $|Nodes(\mathtt{S_1},I_0'',Q)|=|Nodes(\mathtt{S_1},I_0,P)|+|Nodes(\mathtt{S_1},I_0',P')|\geqslant 2(n-2f)>n-f$.
So we should have $|F|\leqslant f-1$ during $I_0''$, which in turn indicates that $|Nodes(\mathtt{S_1},I_0'',Q)|\geqslant 2(n-2f+1)$ holds.
Iteratively, $|Nodes(\mathtt{S_1},I_0'',Q)|>n$ should be satisfied.
So no such $t_2'$ exists and thus the conclusion holds.
\end{IEEEproof}

Then, with these coarsely coordinated engagements, the scheduled next pulsing instants of all bunnies can be adjusted into a short time interval.
And this result would be manifested in the next pulses of the bunnies.

\begin{lemma}
\label{lemma_coordinated}
If $e(\mathtt{D_1},I,Q)\in\mathbf{E_{gg}}$ and $e(\mathtt{L_1},I',Q)=\emptyset$ hold for some $I=[t_0,t_1+\delta_1+2d)$ and $I'=[t_1,t_2)$ with $|I|\leqslant\epsilon$, $|I'|\geqslant \epsilon_2+\delta_1+\delta_2+4d$, $\delta_2\geqslant\vartheta d$ and $T\geqslant (\delta_0+\delta_1+\varepsilon_1+\vartheta(|I|+2\delta_0+9d))/(1-\rho)$, then $e(\mathtt{P}\cup\mathtt{L_1}\cup\mathtt{D_1},[t_1+\delta_1+2d,I_1+\delta_0+2d],Q)=e(\mathtt{P_1},I_1,Q)\in\mathbf{E_{gg}}$ hold with $I_1=I+T+[-\vartheta^{-1}(\rho T+\delta_1+\varepsilon_1),d)$.
\end{lemma}
\begin{IEEEproof}
Denoting $t_q=Times(\mathtt{D_1},I,q)_{-1}$ for every $q\in Q$, by definition we have $e(\mathtt{D_1},(t_q,t_1+\delta_1+2d),q)=\emptyset$.
As $e(\mathtt{D_1},I,Q)\in\mathbf{E_{gg}}$, we have $t_q\in I$ and $\forall q_1,q_2\in Q:|t_{q_1}-t_{q_2}|\leqslant |I|\leqslant \epsilon$.
As $e(\mathtt{L_1},I',Q)=\emptyset$ and $|I'|>\delta_1+2d$, we have $e(\mathtt{L_1}\cup\mathtt{D_1},(t_q,t_2),q)=\emptyset$.
And as $t_2\geqslant t_q+\delta_2+d$, with the line \ref{code_line_engage_k} of Fig.~\ref{algo:bunny-partial} and the higher preemptive priority of the engagement, we have $e(\mathtt{K_1},t_q+[\vartheta^{-1}\delta_2,\delta_2+d),q)\neq \emptyset$.
As $\delta_2\geqslant\vartheta d$, with the lower preemptive priority of the absorption, we have $e(\mathtt{D_0}\cup \mathtt{K_0},[t_q,t_q+d],q)=\emptyset$.
Now denoting $t_q'=Times(\mathtt{P},t_q+d,q)_1$ and $t_q''=Times(\mathtt{P_0},t_q+d,q)_1$, as the $\mathtt{D_0}$ event and the $\mathtt{K_0}$ event can only be generated in responding to the $\mathtt{P}$ event and the $\mathtt{P_0}$ event respectively, we have $e(\mathtt{D_0},[t_q,t_q'),q)=e(\mathtt{K_0},[t_q,t_q''),q)=\emptyset$.
Now assuming $q_0\in Q$ and $t_{q_0}''=\min_{q\in Q}\{t_q''\}$, we have $\forall q\in Q:e(\mathtt{K_0},[t_q,t_{q_0}''),q)=\emptyset$ and thus $\forall q\in Q, t\in [t_q+\delta_2+d,t_{q_0}''): k_\mathtt{A}^{(q)}(t)\neq 0$.
As the $\mathtt{S_2}$ event can only be generated in $q$ with $k_\mathtt{A}^{(q)}= 0$, we have $e(\mathtt{S_2},[t_q+\delta_2+d,t_{q_0}''),q)=\emptyset$ and thus $e(\mathtt{S_2},[t_1+\delta_1+\delta_2+3d,t_{q_0}''),Q)=\emptyset$.
And as $t_2-t_1=|I'|\geqslant \epsilon_2+\delta_1+\delta_2+4d$, we have $e(\mathtt{S_2},[t_2-\epsilon_2-d,t_{q_0}''),Q)=\emptyset$.
So with the definition of $\Psi_\mathtt{engag}$, we have $e(\mathtt{L_1},[t_2,t_{q_0}''),Q)=\emptyset$ and thus $e(\mathtt{L_1}\cup\mathtt{D_1},(t_q,t_{q_0}''),q)=\emptyset$.
Now for this $q_0$, as $e(\mathtt{D_1},(t_{q_0},t_{q_0}''),{q_0})=\emptyset$, $\tau_{sch}^{({q_0})}$ cannot be set by any engagement during $(t_{q_0},t_{q_0}'')$.
With the line \ref{code_line_engage_adjust} of Fig.~\ref{algo:bunny-partial}, the engagement of $q_0$ generated at $t_{q_0}$ can only reference to the marks observed in the latest $\delta_1+\varepsilon_1+2\vartheta d$ ticks.
So we have $\tau_{sch}^{(q_0)}(t_{q_0})\in \tau_{q_0}(t_{q_0})\oplus[-\delta_1-\varepsilon_1-2\vartheta d:0]\oplus T$ and thus $t_{q_0}'\in t_{q_0}+[\vartheta^{-1}(T-\delta_1-\varepsilon_1)-2d, T+d)$ with $t_{q_0}'\leqslant t_{q_0}''$.
With the line \ref{code_line_absorb_adjust} of Fig.~\ref{algo:bunny-partial}, an absorption of ${q_0}$ can only reference to the marks observed in the latest $2\delta_0+2\vartheta d$ ticks.
So with $K_\mathtt{A}\geqslant2$ we have $\tau_{sch}^{(q_0)}(t_{q_0}'')\in \tau_{sch}^{(q_0)}(t_{q_0}')\oplus(K_\mathtt{A}-1)([-\delta_0-2\vartheta d:\delta_0]\oplus T)$ and thus $t_{q_0}''\geqslant t_{q_0}+\vartheta^{-1}(2T-\delta_0-\delta_1-\varepsilon_1)-4d$.
As $2T\geqslant \delta_0+\delta_1+\varepsilon_1+\vartheta(|I|+T+2\delta_0+9d)$, we have $t_{q_0}''\geqslant t_{q_0}+|I|+T+2\delta_0+5d\geqslant \max_{q\in Q}\{t_{q}+T+2\delta_0+5d\}$ and thus $e(\mathtt{L_1}\cup\mathtt{D_1},(t_q,t_q+T+2\delta_0+5d),q)=\emptyset$ for all $q\in Q$.
So, similar to $q_0$, now for every $q\in Q$ we have $t_{q}'\in t_{q}+[\vartheta^{-1}(T-\delta_1-\varepsilon_1), T+d)\subseteq I+T+[-\vartheta^{-1}(\rho T+\delta_1+\varepsilon_1),d)$ and the conclusion holds.
\end{IEEEproof}

\begin{corollary}
\label{corollary_coordinated}
If $e(\mathtt{L_2},t,Q)\neq \emptyset$, $\delta_2=\vartheta d$, and $T\geqslant T_0= \max\{\\(\delta_0+\vartheta(3\epsilon_1+\epsilon_2+2\delta_0+\delta_1+13d))/(1-\rho), \vartheta(3\epsilon_1+2\epsilon_2+\delta_1+\delta_2+\varrho_1+5d)\}$, then $e(\mathtt{P}\cup\mathtt{L_1}\cup\mathtt{D_1},[t+\epsilon_1+\epsilon_2+\delta_1+3d,I_1+\delta_0+2d),Q)=e(\mathtt{P_1},I_1,Q)\in\mathbf{E_{gg}}$ holds for $I_1=t+T+I_\mathtt{A}$ and $I_\mathtt{A}=[-\vartheta^{-1}\rho T-2\epsilon_1-d,\epsilon_1+\epsilon_2+\delta_1+4d)$.
\end{corollary}
\begin{IEEEproof}
As $e(\mathtt{L_2},t,Q)\neq \emptyset$, by applying Lemma~\ref{lemma_steadfast}, $e(\mathtt{D_1},I,Q)\in\mathbf{E_{gg}}$ and $e(\mathtt{L_1},I',Q)=\emptyset$ hold for $I=t+[\vartheta^{-1}\delta_1-\epsilon_1,\epsilon_1+\epsilon_2+\delta_1+3d)$ and $I'=t+[\epsilon_1+\epsilon_2+d,\vartheta^{-1}T-\varrho_1-2\epsilon_1)$.
Now with $T\geqslant T_0$, we have $|I'|\geqslant \epsilon_2+\delta_1+\delta_2+4d$.
Thus with Lemma~\ref{lemma_coordinated}, the conclusion holds with $|I|=2\epsilon_1+\epsilon_2+3d+\vartheta^{-1}\rho\delta_1$ and $\varepsilon_1=\vartheta(\epsilon_1+d)$.
\end{IEEEproof}

Now, as is deduced from the happy bunny, if $T\gg |I_\mathtt{A}|$, the pulsing events of all bunnies would be coarsely coordinated into the relatively short time interval $I_1$.
Upon this, there could be alternative absorption strategies.
By setting the local time parameter $\delta_0$ as $\vartheta(\epsilon_\mathtt{A}+d)$ with $\epsilon_\mathtt{A}=|I_\mathtt{A}|=3\epsilon_1+\epsilon_2+\delta_1+5d+\vartheta^{-1}\rho T$, the algorithm shown in Fig.~\ref{algo:bunny-partial} takes a relaxed two-stage strategy.
Namely, with the coarsely coordinated pulsing events generated during $I_1$, now every $q\in Q$ can observe the marks of all bunnies in the relaxed local time window of $2\delta_0+2\vartheta d$ ticks.
Then, by iteratively executing fault-tolerant averaging (FTA) in the \emph{AdjustByAllFellows} function with these observations (like the Lynch-Welch algorithm \citep{LynchWelch1984}), the pulsing instants of all bunnies can converge into a desired short time interval, as is shown in the following lemma.

\begin{lemma}
\label{lemma_absorption}
If $\delta_0=\vartheta(\epsilon_\mathtt{A}+d)$, $\delta_2=\vartheta d$, $T \geqslant T_1=(\delta_0+\vartheta(\epsilon_\mathtt{A}+2\delta_0+9d))/(1-\rho)$, $e(\mathtt{P}\cup\mathtt{L_1}\cup\mathtt{D_1},I_{k}+(\delta_0+2d)[-1,1],Q)=e(\mathtt{P}_k,I_{k},Q)\in\mathbf{E_{gg}}$ hold for $k\in[1:K_\mathtt{A}-2]$ with $|I_{k}|\leqslant\epsilon_\mathtt{A}$, then $e(\mathtt{P}\cup\mathtt{L_1}\cup\mathtt{D_1},I_{k+1}+(\delta_0+2d)[-1,1],Q)=e(\mathtt{P}_{k+1},I_{k+1},Q)\in\mathbf{E_{gg}}$ hold for some $I_{k+1}\subseteq I_k+T+[-\vartheta^{-1}\rho T,d)$ with $|I_{k+1}|\leqslant|I_k|/2+d+\rho\vartheta^{-1} T$.
\end{lemma}
\begin{IEEEproof}
As $e(\mathtt{P}_k,I_k,Q)\in\mathbf{E_{gg}}$ and $e(\mathtt{L_1}\cup\mathtt{D_1},I_k+[-\delta_2-d,\delta_0+2d],Q)=\emptyset$, we have $e(\mathtt{D_0},I,Q)\in\mathbf{E_{gg}}$ for $I=I_k+[\vartheta^{-1}\delta_0,\delta_0+2d)$.
And similar to the proof of Lemma~\ref{lemma_coordinated}, denoting $t_q=Times(\mathtt{P}_k,I_k,q)_{-1}$, $t_q'=Times(\mathtt{P},t_q,q)_2$, $t_q''=Times(\mathtt{P_0},t_q,q)_1$ for every $q\in Q$ and $t_{q_0}''=\min_{q\in Q}\{t_q''\}$ for some $q_0\in Q$, we have $e(\mathtt{L_1}\cup\mathtt{D_1},(t_q,t_{q_0}''),q)=\emptyset$.
As $1\leqslant k<K_\mathtt{A}-1$ and $T\geqslant T_1$, we have $\tau_{sch}^{(q_0)}(t_{q_0}'')\in \tau_{q_0}(t_{q_0})\oplus(K_\mathtt{A}-k)([-\delta_0-2\vartheta d:\delta_0]\oplus T)$ and thus $t_{q_0}''\geqslant t_{q_0}+2\vartheta^{-1}(T-\delta_0)-4d$ and $e(\mathtt{L_1}\cup\mathtt{D_1},I_{k}+[-\delta_2-d,T+2\delta_0+5d],Q)=\emptyset$.
Thus, during $I_{k}+[0,T+2\delta_0+5d]$, for all $q\in Q$, $\tau_{sch}^{({q})}$ can only be set in \emph{Absorb} in responding to the $\mathtt{P}$ events and any such \emph{Absorb} would not be preempted by \emph{Engage}.
So as $e(\mathtt{D_0},I,Q)\in\mathbf{E_{gg}}$ and $e(\mathtt{P},I_k+(\delta_0+2d)[-1,1],Q)=e(\mathtt{P}_k,I_k,Q)$ with $|I_k|\leqslant\epsilon_\mathtt{A}\leqslant\vartheta^{-1}\delta_0-d$, a round of (synchronous) approximate agreement is simulated in $e(\mathtt{D_0},I,Q)$.
Concretely, denoting $\psi_{ref.q}=\bar b_0(\psi_q(Times(\mathtt{D_0},I,q)_{-1})[J_{2\delta_0+2\vartheta d}])$ as the referenced trail in $q$, as $Ticks(\psi_{ref.q},Q)\subseteq \tau_q(I_k+[0,d))$, by applying the convergence property of approximate agreement, we have $\forall q_1,q_2\in Q:|\tau_{q_1}^{-1}(\tau_{sch}^{({q_1})}(t_{q_1}+\delta_0+2d)\ominus T)-\tau_{q_2}^{-1}(\tau_{sch}^{({q_2})}(t_{q_2}+\delta_0+2d)\ominus T)|< |I_k|/2+d$.
Thus, as $\tau_{sch}^{(q)}(t_{q}+\delta_0+2d)=\tau_q(t_q')$, we have $\forall q_1,q_2\in Q:|t_{q_1}'-t_{q_2}'|< |I_k|/2+d+\rho\vartheta^{-1} T$.
Besides, by applying the closure property of approximate agreement, we have $\forall q\in Q:\tau_{q}^{-1}(\tau_{sch}^{({q})}(t_{q}+\delta_0+2d)\ominus T)\in I_k+[0,d)$ and thus $\forall q\in Q:\tau_{q}^{-1}(\tau_{sch}^{({q})}(t_{q}+\delta_0+2d))\in I_k+[0,d)+T[\vartheta^{-1},1]$.
Also, as $k_\mathtt{A}^{(q)}(t_q)=k$, we have $k_\mathtt{A}^{(q)}(t_q')=k+1$ by executing the \emph{Absorb} during $[t_q,t_q']\subseteq I_{k}+[0,T+2\delta_0+5d]$ and thus the conclusion holds.
\end{IEEEproof}

\begin{corollary}
\label{corollary_absorption}
If the premise of Lemma~\ref{lemma_absorption} holds for $k=1$, then for every $1< k<K_\mathtt{A}$ there exists $I_{k}\subseteq I_1+(k-1)(T+[-\vartheta^{-1}\rho T,d))$ with $|I_{k}|\leqslant 2^{1-k}\epsilon_\mathtt{A}+2(d+\rho\vartheta^{-1} T)$ satisfying $e(\mathtt{P}\cup\mathtt{L_1}\cup\mathtt{D_1},I_{k}+(\delta_0+2d)[-1,1],Q)=e(\mathtt{P_{k}},I_{k},Q)\in\mathbf{E_{gg}}$.
\end{corollary}
\begin{IEEEproof}
By iteratively applying Lemma~\ref{lemma_absorption}, the premise of Lemma~\ref{lemma_absorption} holds for every $k\in[2:K_\mathtt{A}-1]$.
As $I_{k}\subseteq I_{k-1}+T+[-\vartheta^{-1}\rho T,d)$, we have $I_{k}\subseteq I_1+(k-1)(T+[-\vartheta^{-1}\rho T,d))$.
As $|I_1|\leqslant \epsilon_\mathtt{A}$, we have $|I_{k}|\leqslant 2^{1-k}\epsilon_\mathtt{A}+2(1-2^{1-k})(d+\rho\vartheta^{-1} T)$ and thus the conclusion holds.
\end{IEEEproof}

So, by properly setting $\delta_2$, $\delta_0$, and $T$, a happy bunny would definitely know all bunnies can be absorbed by the $K_\mathtt{A}$-cycle two-stage absorption process if this process is not disturbed by any preemption of a new engagement or appearance of \emph{Hoppelpopp}.
Now for this two-stage absorption process being followed by a new one, every $q\in Q$ should make a new engagement at the end of the current absorption process.
For this, the local accuracy threshold $\varrho_1$ should be properly set to allow every $q\in Q$ to tag its mark with $\mathtt{GOOD}$ before the end of the current absorption process.
Namely, with the bounded duration between the $(k-1)$th and the $k$th pulsing instants in $q$ after the last engagement, $\varrho_1$ should be set sufficiently large to cover all possible cases for $k\geqslant K_\varrho$ with some $K_\varrho< K_\mathtt{A}$.
Besides, $\varrho_1$ should also be set sufficiently small for the engagements being coarsely coordinated (required in Lemma~\ref{lemma_coordinated}).
Now by satisfying both the requirements, every $q\in Q$ would be good since $k_\mathtt{A}^{(q)}\geqslant K_\varrho$, as is shown in the following lemma.

\begin{lemma}
\label{lemma_good}
If $\varrho_1= \rho T+\vartheta(\epsilon_\varrho+2d)$ with $\epsilon_\varrho=2^{2-K_\varrho}\epsilon_\mathtt{A}+2(d+\rho\vartheta^{-1} T)$, $K_\varrho<K_\mathtt{A}$ and the premise of Corollary~\ref{corollary_absorption} hold, then $e(\mathtt{S_1},I_{k}+[0,d),Q)\in\mathbf{E_{gg}}$ holds for every $K_\varrho\leqslant k< K_\mathtt{A}$.
\end{lemma}
\begin{IEEEproof}
Firstly, with Corollary~\ref{corollary_absorption} and Lemma~\ref{lemma_absorption}, we have $|I_{K_\varrho}|\leqslant|I_{K_\varrho-1}|\leqslant \epsilon_\varrho$ and $I_{K_\varrho}\subseteq I_{K_\varrho-1}+T+[-\rho\vartheta^{-1} T,d)$.
Thus, denoting $t_q^{k}=Times(\mathtt{P_{k}},I_{k},q)_{-1}$, we have $t_q^{k}-t_q^{k-1}\in T+[-\epsilon_\varrho,\\\epsilon_\varrho]+[-\rho\vartheta^{-1} T,d)$ for every $q\in Q$ and $K_\varrho\leqslant k< K_\mathtt{A}$.
So we have $T-(\epsilon_\varrho+\rho\vartheta^{-1} T)\leqslant\tau_q(t_q^{k})\ominus\tau_q(t_q^{k-1})\leqslant \vartheta(T+\epsilon_\varrho+d)=T+\rho T+\vartheta(\epsilon_\varrho+d)$.
As $\varrho_1\geqslant \max\{\epsilon_\varrho+\rho\vartheta^{-1} T,\rho T+\vartheta(\epsilon_\varrho+d)\}+\vartheta d$, we have $\psi_q(t_q^{k})\in\Psi_\mathtt{good}$ and thus $e(\mathtt{S_1},t_q^{k}+[0,d),q)\neq \emptyset $ holds for every $q\in Q$ and $K_\varrho\leqslant k< K_\mathtt{A}$.
As $t_q^{k}\in I_{k}$, $e(\mathtt{S_1},I_{k}+[0,d),Q)\in\mathbf{E_{gg}}$ holds for every $K_\varrho\leqslant k< K_\mathtt{A}$.
\end{IEEEproof}

With this, the next step is that every $q\in Q$ should tag its mark with $\mathtt{GOOD}$ and $\mathtt{BEST}$ in responding to the new $\mathtt{P_0}$ event, by which the new two-stage absorption process should in turn be initiated.
For this, firstly, the pulsing instants of all $q\in Q$ should be sufficiently closed to each other for each $q$ being able to find a shortcut trail before its new $\mathtt{P_0}$ event.
This can be satisfied by setting $\epsilon_0$ and $K_\mathtt{A}$ being sufficiently large.
And secondly, the time parameter $\delta_1$ should be properly set to avoid the new $\mathtt{P_0}$ event being prevented in $q$ by a too-early engagement which reset the scheduling timer $\tau_{sch}$ before it is expired (i.e., being reached by the current local time $\tau_q$).
Now with the sufficiently large $\epsilon_0$ and $K_\mathtt{A}$, by setting $\delta_1\geqslant\varepsilon_0$, every $q\in Q$ would participate in a new desired absorption process, as is shown in the following lemma.

\begin{lemma}
\label{lemma_initiation}
If $\epsilon_0> 2(d+\rho\vartheta^{-1} T)$, $\delta_1\geqslant\varepsilon_0$, $K_\mathtt{A}\geqslant K_\epsilon+1$ with $K_\epsilon=1+\lceil \log_2 \epsilon_\mathtt{A}/(\epsilon_0-2(d+\rho\vartheta^{-1} T))\rceil$ and the premise of Lemma~\ref{lemma_good} hold, then $e(\mathtt{S_2},I'+(\delta_1+\varepsilon_1+2\vartheta d)[-1,1],Q)=e(\mathtt{S_2},I',Q)\in\mathbf{E_{gg}}$ and $\{e(\mathtt{L_2},I',Q),e(\mathtt{L_1},I',Q)\}\subseteq\mathbf{E_{gg}}$ hold for some $I'=I_{0}'+[0,d)$ with $I_{0}'\subseteq I_1+(K_\mathtt{A}-1)(T+[-\vartheta^{-1}\rho T,\\d))$ and $|I_{0}'|\leqslant\epsilon_0$.
\end{lemma}
\begin{IEEEproof}
Firstly, with Corollary~\ref{corollary_absorption} and Lemma~\ref{lemma_good}, for all $K_\varrho\leqslant k< K_\mathtt{A}$, $e(\mathtt{S_1},I_{k}+[0,d),Q)\in\mathbf{E_{gg}}$ holds with $|I_{k}|\leqslant 2^{1-k}\epsilon_\mathtt{A}+2(d+\rho\vartheta^{-1} T)$.
As $K_\mathtt{A}-1\geqslant K_\epsilon$, we have $|I_{K_\mathtt{A}-1}|\leqslant \epsilon_0$.
Thus, by the definition of $\Psi_\mathtt{short}$, every $q\in Q$ would see a shortcut trail and thus set $\mathtt{isBest}^{(q)}$ being true during $I_{K_\mathtt{A}-1}+[0,d)$.
And as $\delta_1\geqslant\varepsilon_0=\vartheta(\epsilon_0+d)$, every $q\in Q$ would distribute a mark message with value $\{\mathtt{GOOD,BEST}\}$ in responding to $e(\mathtt{P_0},I_0',q)$, where $I_{0}'\subseteq I_{K_\mathtt{A}-1}+T+[-\rho\vartheta^{-1} T,d)\subseteq I_1+(K_\mathtt{A}-1)(T+[-\vartheta^{-1}\rho T,d))$ with $|I_{0}'|\leqslant \epsilon_0$.
Thus, similar to the proof of Lemma~\ref{lemma_absorption}, we have $e(\mathtt{S_2},I'+(\delta_1+\varepsilon_1+2\vartheta d)[-1,1],Q)=e(\mathtt{S_2},I',Q)\in\mathbf{E_{gg}}$.
And as the total delays of the mark messages are all bounded within $d$, we also have $Nodes(\mathtt{L_2},I',Q)=Nodes(\mathtt{L_1},I',Q)=Q$.
\end{IEEEproof}

Now together with Lemma~\ref{lemma_good}, Lemma~\ref{lemma_initiation}, Lemma~\ref{lemma_coordinated}, Corollary~\ref{corollary_coordinated} and Lemma~\ref{lemma_absorption}, for simplicity we can set $K_\varrho=K_\epsilon+1$, $K_\epsilon=1+\lceil \log_2 \epsilon_\mathtt{A}/(\epsilon_0-2(d+\rho\vartheta^{-1} T))\rceil$, $\varrho_1= \rho T+\vartheta(\epsilon_\varrho+d)+\vartheta d$, $\epsilon_\varrho=2^{2-K_\varrho}\epsilon_\mathtt{A}+2(d+\rho\vartheta^{-1} T)$, $\delta_2=\vartheta d$, $\delta_1=\varepsilon_0=\vartheta(\epsilon_0+d)$, $\delta_0=\vartheta(\epsilon_\mathtt{A}+d)$, $\epsilon_\mathtt{A}=3\epsilon_1+\epsilon_2+\delta_1+5d+\vartheta^{-1}\rho T$ for some $\epsilon_0\geqslant 2(d+\rho\vartheta^{-1} T)$, $K_\mathtt{A}>K_\varrho$ and $T\geqslant\vartheta(3\delta_0+\epsilon_\mathtt{A}+9d)/(1-\rho)$ in solving the required parameters.
And as $0\leqslant\rho\ll 1$, the required inequalities can be easily satisfied.
So we take these settings as the default in the rest of this paper.
With this, the last engagements of every $q\in Q$ in responding to the $e(\mathtt{L_1},I',Q)$ events given in Lemma~\ref{lemma_initiation} would bring the paces of all bunnies into the desired synchronization.
Then, a new absorption process would follow the last absorption process, with which the synchronization precision and accuracy having been reached can be inherited.
Further, as each $q\in Q$ would be happy at some instant $t_0\in I'$, $q$ would be always happy since $t_0$, providing that no \emph{Hoppelpopp} appears since $I'$.
On the whole, denoting $T^-=\vartheta^{-1}T-2\epsilon_0$, $T^+=T+2\epsilon_0$, $\phi_0^-=K_\mathtt{A} T^-$ and $\phi_0^+=K_\mathtt{A} T^+$, we have the following result in this section.

\begin{lemma}
\label{lemma_absorption_process}
If $e(\mathtt{H},[t-\epsilon_2,t_1],Q)=\emptyset$ and $e(\mathtt{L_2},t,Q)\neq \emptyset$ hold with $t_1-t>j\Delta_\mathtt{A}$, $\Delta_\mathtt{A}=\epsilon_\mathtt{A}+K_\mathtt{A}(T+d)+\epsilon_0+d$ and $j\geqslant 1$, then $e(I^*,Q)\in\mathbf{E}_{\epsilon_0,T^-,T^+}^{|Q|,1+(j'-1)K_\mathtt{A}}$, $\bar s_{\mathtt{L_2}}(e(\mathtt{L_2},I^*,Q))\in \mathbf{E}_{\epsilon_0+d,\phi_0^-,\phi_0^+}^{|Q|,j'}$ and $(\mathtt{L_2},I^*,Q)_1\leqslant t+\Delta_\mathtt{A}$ hold with some $I^*=[t^*,t_1^*]$, $t^*\in [t,t+\Delta_\mathtt{A}]$, $t_1^*\in [t_1-\epsilon_0-d,t_1]$ and $j'\geqslant j$.
\end{lemma}
\begin{IEEEproof}
As $e(\mathtt{H},[t-\epsilon_2,t_1],Q)=\emptyset$, no event of $Q$ is generated in responding to the $\mathtt{H}$ event during $[t-\varepsilon_1,t_1]$.
So $e(\mathtt{H},I,Q)=\emptyset$ holds for all $I$ referred in the former lemmata.
As $e(\mathtt{L_2},t,Q)\neq \emptyset$, with Corollary~\ref{corollary_coordinated} and Lemma~\ref{lemma_initiation} we have $e(\mathtt{S_2},I'+(\delta_1+\varepsilon_1+2\vartheta d)[-1,1],Q)=e(\mathtt{S_2},I',Q)\in\mathbf{E_{gg}}$, $e(\mathtt{L_2},I',Q)\in \mathbf{E_{gg}}$ and $e(\mathtt{L_1},I',Q)\in \mathbf{E_{gg}}$ with $I'=I_{0}'+[0,d)$, $|I_{0}'|\leqslant \epsilon_0$, $I_{0}'\subseteq I_1+(K_\mathtt{A}-1)(T+[-\vartheta^{-1}\rho T,d))$, $I_1=t+T+I_\mathtt{A}$, $I_\mathtt{A}\subseteq \epsilon_\mathtt{A}(-1,1)$.
Now as $\delta_1\geqslant\varepsilon_0$ and $|I'|<\epsilon_0+d$, denoting $t_q=Times(\mathtt{D_1},I'+[\vartheta^{-1}\delta_1,\delta_1+2d),q)_{-1}$, the events $e(\mathtt{D_1},t_q,q)$ of all $q\in Q$ simulate a round of the approximate agreement.
Thus, similar to the proof of Lemma~\ref{lemma_absorption}, we have $e(\mathtt{P}\cup\mathtt{L_1}\cup\mathtt{D_1},I_{1}'+(\delta_0+2d)[-1,1],Q)=e(\mathtt{P_1},I_{1}',Q)\in\mathbf{E_{gg}}$ holds with $|I_{1}'|\leqslant\epsilon_0$.
So the conclusion holds by iteratively applying Lemma~\ref{lemma_initiation} and Lemma~\ref{lemma_absorption}.
\end{IEEEproof}

Now it can be explained why an engaged bunny would not claim itself to be the best.
It is for the constant-time two-stage absorption process which mainly consists of one engagement and several following absorptions.
The main advantage of the two-stage process is that the coarser initial coordination (with the engagement) and the finer iterative convergence (with the following absorptions) can be temporally decoupled.
In principle, this relaxed two-stage strategy gives a chance for the bunnies to take a wider time window in the following absorptions in embracing all possible marks from the initially engaged bunnies to simulate the desired approximate agreements.
And this can hardly be accomplished with a single-stage strategy where the bunnies should both take an eye on the self-claimed best bunnies and the might-be fellows.
In this sense, being \emph{good} is sometimes better than being the \emph{best} in reaching P-SSBPS.
Now with the two-stage absorption process during which no $\mathtt{H}$ event is generated, the system can be synchronized in the eyes of a happy bunny.
But there might be no happy bunny or the undesired $\mathtt{H}$ event might be generated.
And these cases are handled in the next two sections.

%
%

\section{The Emergency Process}
\label{sec:emergency}
In the last paragraph of Section~\ref{subsec:Hoppelpoppand}, the readers might wonder why \emph{Hoppelpopp} cannot always be lucky.
To ascertain this, it should be understood that the required \emph{Hoppelpopp} cannot be realized by any single node in the distributed fault-tolerant system.
Instead, in typical scenarios, \emph{Hoppelpopp} can only be realized by the $|Q|$ bunnies and $|F|$ badgers.
In this situation, to make \emph{Hoppelpopp} being always lucky, the bunnies should run a self-stabilized DBA.
But to perform such a self-stabilized DBA, as the system can be with an arbitrary initial state, it requires at least a linear time $O(f)$ for reaching deterministic stabilization.
Just before the stabilization of the underlying DBA protocol, we cannot expect the \emph{Hoppelpopp} be always lucky.
Further, for faster stabilization, one might expect that the linear-time DBA could be avoided, which means that the expected public appearance of \emph{Hoppelpopp} could be probabilistic even when the underlying (randomized) BA protocol is stabilized.
Generally, when the underlying protocols are all stabilized since $\Delta_\mathtt{c}$, the \emph{Hoppelpopp} should promise that in every $E^\chi$
\begin{eqnarray}
\label{eq_promise_call1}
\emph{Probabilistic-Liveness}:\forall t\geqslant \Delta_\mathtt{c}:\forall t'\geqslant t+\Delta_0 :\nonumber\\
Prob(e(\mathtt{H}\cup\mathtt{L_2},[t,t'],Q)\neq\emptyset)\geqslant\eta_0\\
\label{eq_promise_call0}
\emph{Peaceability}: \forall t\geqslant \Delta_\mathtt{c}: \forall t'\geqslant t+\Delta_1:\forall t_1\in [t,t']:\nonumber\\
 \forall q\in Q: \psi_q(t_1)\in \Psi_\mathtt{happy} \to e(\mathtt{G},[t+\Delta_1,t'],Q)=\emptyset\\
\label{eq_promise_come1}
\emph{Probabilistic-GoodLuck}: \forall t\in (\mathtt{H},[\Delta_\mathtt{c},+\infty),Q):\nonumber\\
 Prob(\exists I\in \mathbf{I_\mathtt{H}}: t\in I )\geqslant\eta_0\\
\label{eq_promise_come0}
\emph{Probabilistic-Separation}: \forall I\in \mathbf{I_\mathtt{H}}\cap [\Delta_\mathtt{c},+\infty):\nonumber\\
Prob(\exists t'\in I+\Delta_2: e(\mathtt{H},[t',t'+\Delta_3],Q)=\emptyset)\geqslant\eta_0
\end{eqnarray}
holds, where $\mathbf{I_\mathtt{H}}=\{I\mid e(\mathtt{H},I,Q)\in\mathbf{E_{gg}} \land |I|\leqslant \epsilon_\mathtt{H}\}$, $\epsilon_\mathtt{H}$ is a fixed upper-bound of the consistent appearance errors, $Prob(x)$ is the probability of $x$ being true and $\eta_0\in (0,1]$ is a fixed probability.
Besides, for the P-SSBPS solutions, the durations $\Delta_\mathtt{c}$, $\Delta_0$, $\Delta_1$, $\Delta_2$ and $\Delta_3$ should be finite numbers satisfying some desired relations, as is detailed in the analysis.

\subsection{Realization of Hoppelpopp}
\label{subsec:realize-Hoppelpopp}
To fulfill these promises, firstly, we employ the self-stabilizing \emph{Initiator-Accept} primitive (also called \emph{I-Accept}, denoted as $\mathcal P_{IA}$) proposed in \cite{Daliot2006Agreement} to initiate the semi-synchronous BA process.
This BA process, denoted as $e(\mathtt{B},I,Q)$, is used to simulate a $k$-round ($k\leqslant K_\mathtt{B}$) synchronous BA protocol $\mathcal{B}$ with $|I|\leqslant \Delta_\mathtt{B}$ and $\mathtt{B}\subset\mathtt{G}$ being the event types exclusively used in $\mathcal{B}$.
As $\mathcal{B}$ should be terminated in bounded rounds, in the most natural way we can directly employ the classical fixed-round immediate DBA protocol \citep{Berman1989optimalconsensus} or early-stopping eventual DBA protocol \citep{Toueg1987Fast,Dolev1990Early,Berman1992early} as the underlying $\mathcal{B}$ protocol.
With this, as is shown in \cite{Daliot2006Agreement}, the $\mathcal P_{IA}$ primitive and the BA process can be self-stabilized in linear-time deterministically (i.e., with probability $1$).
Further, by combining DBA and RBA protocols such as in \cite{Goldreich1990bestbothworlds}, the common-coin-based randomized protocols \citep{FM1989,FM1997} can be integrated with early-stopping DBA protocols to construct some \emph{hybrid} $\mathcal{B}$ protocols that are expected to terminate earlier.
With this, it is expected that the P-SSBPS system could also be somewhat stabilized further earlier in some cases.

Now, as our main interest here is to realize the so-called \emph{Hoppelpopp}, the analyses of the related existing algorithms are not re-investigated here.
Instead, in trying to \emph{stand on the tall shoulders} \citep{Berman1989optimalconsensus,Toueg1987Fast,Daliot2006Agreement,FM1989,FM1997,Goldreich1990bestbothworlds}, the core properties of the underlying algorithms are sketched here in words of processes.
Firstly, by following the \emph{Sending Validity Criteria} required in \cite{Daliot2006Agreement}, the self-stabilized $\mathcal P_{IA}$ primitive satisfies
\begin{eqnarray}
\label{eq_primitive_IA_Correctness1}
\emph{Extended Correctness 1}:\nonumber\\
G,q\in Q\land e(\mathtt{G_2},t,q)\neq\emptyset\to\tau_q(t)\ominus\tau_q^G(t)\leqslant 4\vartheta d\\
\label{eq_primitive_IA_Correctness2}
\emph{Extended Correctness 2}:\nonumber\\
|I|< d \land |Nodes(\mathtt{G_1},I,Q)|> f \to\nonumber\\
 e(\mathtt{G_2},I+[0,3d],Q)\in\mathbf{E_{gg}}\\
\label{eq_primitive_IA_Unforgeability}
\emph{Extended Unforgeability}:\nonumber\\
t'\geqslant t+\Delta_\mathtt{v} \land e(\mathtt{G_1},[t,t'],Q)=\emptyset \to \nonumber\\
e(\mathtt{G_2},[t+\Delta_\mathtt{v},t'],Q)=\emptyset\\
\label{eq_primitive_IA_Uniqueness}
\emph{Extended Uniqueness}:\nonumber\\
e(\mathtt{G_2},t_1,q_1)\neq\emptyset \land e(\mathtt{G_2},t_2,q_2)\neq\emptyset \to \nonumber\\
|\tau_{q_1}^{-1}(\tau_{q_1}^G(t_1))-\tau_{q_2}^{-1}(\tau_{q_2}^G(t_2))|\notin [6d,2\Delta_\mathtt{rmv}-3d]\\
\label{eq_primitive_IA_Relay}
\emph{Extended $\Delta_\mathtt{agr}$-\emph{Relay}}:\nonumber\\
t\leqslant \tau_{q}^{-1}(\tau_{q}^G(t))+\Delta_\mathtt{agr} \land e(\mathtt{G_2},t,q)\neq\emptyset \to \nonumber\\
e(\mathtt{G_1},[\tau_{q}^{-1}(\tau_{q}^G(t)),t],Q)\neq\emptyset \land \nonumber\\
\exists I: t\in I \land |I|\leqslant 2d \land e(\mathtt{G_2},I,Q)\in\mathbf{E_{gg}}
\end{eqnarray}
for a specific \emph{General} $G$ with a stabilization time no more than $\Delta_\mathtt{stb}$ \citep{Daliot2006Agreement,DolevPulseBoundedDelay2007}.
Here, $\mathtt{G_0},\mathtt{G_1},\mathtt{G_2}\in \mathtt{G}$ are respectively the event types of distributing, responding and I-accepting the $IA$ messages of $G$.
And according to \cite{Daliot2006Agreement}, the instant of generating the $\mathtt{G_0}$ event in the specific \emph{General} $G$ is estimated as the local time $\tau_q^G$ in node $q$.
Also, the time parameters $\Delta_\mathtt{stb}=2(20d+4\Delta_\mathtt{rmv})$, $\Delta_\mathtt{v}=2\Delta_\mathtt{rmv}+15d$ and $\Delta_\mathtt{rmv}=\Delta_\mathtt{agr}+13d$ all take values following \cite{Daliot2006Agreement} but with $\Delta_\mathtt{agr}=\Delta_\mathtt{B}$.
For simplicity, we assume the $IA$ messages are all with the same value in the basic analysis.

Then, by differentiating the various $\mathcal{B}$ protocols, the P-SSBPS solutions can be divided into different categories.
For example, when $\mathcal{B}$ is realized as some fixed-round immediate DBA protocols (like \cite{Berman1989optimalconsensus}), the corresponding P-SSBPS solutions are referred to as the simple basic solutions.
And for $\mathcal{B}$ being realized as some early-stopping eventual DBA protocols for some earlier agreement in some non-worst cases, the corresponding P-SSBPS solutions are referred to as the early-synchronizing basic solutions.
Further, the solutions are referred to as the (early-synchronizing) hybrid ones if $\mathcal{B}$ is realized as some hybrid $\mathcal{B}$ protocols.
In all cases, the adversaries are assumed according to the underlying protocols.
For example, in discussing the basic solutions, we assume the \emph{omniscient} but static adversary who can know everything of the system during all possible executions (even including the events that have not been generated at the current time).
In this case, all messages communicated between the nonfaulty nodes are open to the \emph{rushing} adversary \citep{King2011Breaking}.
Further, the \emph{omniscient} adversary can even know all local events generated in the nonfaulty nodes before the related messages being dispatched to the communication network.
In this case, there cannot be any shared secret in the system.
And when we employ the common-coin-based RBA protocols \citep{FM1989,FM1997}, as is required in these underlying protocols, we assume private communication channels and the adversary is not able to know the shared secrets of the nonfaulty nodes before the corresponding \emph{Graded-Recover} stage in each iteration.
Nevertheless, we take the non-cryptographic assumption \citep{BenOr1988NonCryptographic,King2011Breaking,Dolev2014PulseGeneration} in this setting, i.e., the computational power of the adversary is unlimited.
And in a broader perspective, all these solutions are referred to as conservative solutions.
Later in this paper, we would also further extend the system with some cryptographic assumptions or other more benign settings.
In all cases, the desired $\mathcal{B}$ protocol should support the following properties
\begin{eqnarray}
\label{eq_B_Correctness1}
\emph{Deterministic Correctness 0}:~ e(\mathtt{G_3^{0}},t,Q)\in \mathbf{E_{gg}}\to \nonumber\\
Prob(\exists I\subseteq t+[0,\Delta_{\mathtt{B}}]:|I|\leqslant \Delta_{\mathtt{round}} \land e(\mathtt{G_4^{0}},I,Q))=1~\\
\label{eq_B_Correctness2}
\emph{Relaxed Correctness 1}:~ e(\mathtt{G_3^{1}},t,Q)\in \mathbf{E_{gg}}\to \nonumber\\
Prob(\exists I\subseteq t+[0,\Delta_{\mathtt{B}}]:|I|\leqslant \Delta_{\mathtt{round}} \land e(\mathtt{G_4^{1}},I,Q))\geqslant \eta_0\\
\label{eq_B_Agreement}
\emph{Relaxed Agreement 1}:~ e(\mathtt{G_4^{1}},t,Q)\neq \emptyset \to \nonumber\\
Prob(\exists I\subseteq t+[0,\Delta_{\mathtt{B}}]:|I|\leqslant \Delta_{\mathtt{round}} \land e(\mathtt{G_4^{1}},I,Q))\geqslant \eta_0
\end{eqnarray}
when it is ideally executed without any interference, timing error nor local clock drift.
Here, $\Delta_{\mathtt{round}}$ is the ideal duration of a synchronous round and $\eta_0\in(0,1]$ is a fixed positive probability.
We use the event-types $\mathtt{G_3^{v}}$ and $\mathtt{G_4^{v}}$ to respectively represent a $\mathcal{B}$ instance being called with the inputs $\mathtt v$ and outputting $\mathtt v$ in a nonfaulty node for $\mathtt v\in \{0,1\}$.
And for convenience we denote $\mathtt{G_3}=\mathtt{G_3^{0}}\cup \mathtt{G_3^{1}}$ and $\mathtt{G_4}=\mathtt{G_4^{0}}\cup \mathtt{G_4^{1}}$.
Obviously, the classical immediate DBA can satisfy (\ref{eq_B_Correctness1}), (\ref{eq_B_Correctness2}), and (\ref{eq_B_Agreement}) with $\eta_0=1$.
And by allowing the $\mathcal{B}$ protocol being adjacent-round terminable, the finite-round eventual BA that terminates in the adjacent two synchronous round can also satisfy (\ref{eq_B_Correctness1}), (\ref{eq_B_Correctness2}), and (\ref{eq_B_Agreement}) with $\eta_0=1$.

With these, the basic $\mathtt{Hoppelpopp}$ algorithm is constructed as in Fig.~\ref{fig:algo_emergency}.
Firstly, in Fig.~\ref{algo:hop-init}, each $q\in Q$ would act as a \emph{General} to distribute an $IA$ message in executing \emph{Help} if no such message is distributed in $q$ for a sufficiently long duration.
Then, by receiving such a message from $p\in V$, an unhappy $q$ would \emph{explicitly invoke} (see \cite{Daliot2006Agreement}) the $IA$ primitive exclusively running for the \emph{General} $p$, providing that no $\mathtt{H}$ event is generated in the latest $\Delta_\mathtt{relax}$ ticks in $q$.
Here,  a local timer $\tau_x$ in $q$ would be automatically closed if no $\text{set}(\tau_x,\Delta_x)$ is executed during the latest $\Delta_x$ ticks in $q$.

\alglanguage{pseudocode}
\algrenewcommand{\algorithmiccomment}[1]{~//#1}
\begin{figure}[htbp]
\centering
\begin{subfigure}{.49\textwidth}
\begin{algorithmic}[1]
\Statex \underline{\emph{Help}}:
    \If {$\text{closed}(\tau_\mathtt{v})$}
        \State  distribute $(\mathtt{Initiator},q)$;\Comment{$\mathtt{G_0}$} \label{code_line_hop_send}
        \State  $\text{set}(\tau_\mathtt{v},\Delta_\mathtt{v})$;
    \EndIf
\Statex \underline{receive $(\mathtt{Initiator},p)$}: \label{code_line_hop_init}
    \If {not isHappy and $\text{closed}(\tau_\mathtt{relax})$} \label{code_line_hop_init_do}
        \State call \emph{I-Accept}($p$);\Comment{$\mathtt{G_1}$}
    \EndIf
\end{algorithmic}
\caption{The initiation of \emph{Hoppelpopp}}
\label{algo:hop-init}
\end{subfigure}
\begin{subfigure}{.49\textwidth}
\begin{algorithmic}[1]
\Statex \underline{when \emph{I-Accepted}$(p,\tau_q^{p})$}:\Comment{$\mathtt{G_2}$} \label{code_line_hop_IA}
        \State  $\tau_{0}:=\tau_q\ominus\tau_q^{p}$;
        \If{$\tau_{0} \leqslant 6\vartheta d$} \label{code_line_hop_con}
            \State  $c_p:=(\tau_{0} \leqslant 4\vartheta d \land \text{closed}(\tau_\mathtt{relax}))$;
            \State  call $\mathcal{B}_{p}'(c_p)$;\Comment{$\mathtt{G_3}$}
        \Else \Comment{actually do nothing}
            \State  \quad\Comment{call $\mathcal{B}_{p}'(0)$ silently;}
        \EndIf
\end{algorithmic}
\caption{The coordination of $\mathcal{B}'$}
\label{algo:hop-think}
\end{subfigure}
\begin{subfigure}{.49\textwidth}
\begin{algorithmic}[1]
\Statex \underline{$\mathcal{B}_{p}'(c_p)$}:\Comment{can be self-preempted} \label{code_line_hop_simu}
    \State  $k_\mathtt{B}:=0$;
            ~ wait $\delta_\mathtt{B}$ ticks;
    \While {$\mathcal{B}(c_p)$ not return and $k_\mathtt{B}< K_\mathtt{B}$}
        \State  wait $\delta_\mathtt{B}$ ticks;
                ~$k_\mathtt{B}:=k_\mathtt{B}+1$;
        \State  run the $k_\mathtt{B}$th round of $\mathcal{B}(c_p)$;
    \EndWhile
    \If{$\mathcal{B}(c_p)$ returns $c_p=1$}\Comment{$\mathtt{G_4}$}
        \State generate an $\mathtt{H}$ event;
        \State $\text{set}(\tau_\mathtt{relax},\Delta_\mathtt{relax})$;\Comment{$\mathtt{G_5}$} \label{code_line_hop_relax}
    \EndIf
\end{algorithmic}
\caption{The simulation of $\mathcal{B}$}
\label{algo:hop-simu}
\end{subfigure}
\caption{The basic $\mathtt{Hoppelpopp}$ algorithm for each $q\in Q$.}
\label{fig:algo_emergency}
\end{figure}

Then, when the $IA$ message of any \emph{General} $p$ is accepted in $q$, the desired properties of $\mathcal P_{IA}$ should be applied in simulating the following $\mathcal{B}$ protocol.
However, despite the desired properties, an undesired property of $\mathcal P_{IA}$ is that there could be arbitrarily postponed relays other than the time-bounded ones described in (\ref{eq_primitive_IA_Relay}).
That is, there might be $t> \tau_{q}^{-1}(\tau_{q}^G(t))+\Delta_\mathtt{agr}$ in generating $e(\mathtt{G_2},t,q)$ and thus the premise of the $\Delta_\mathtt{agr}$-\emph{Relay} property cannot always be satisfied.
So the problem here is how to integrate the possible inconsistent $\mathtt{G_2}$ events and the following dense-time simulation of the $\mathcal{B}$ protocol.
For this, as the underlying $\mathcal{B}$ protocols might need to exchange several rounds of extra messages (for example the common-coin-based BA), the \emph{too-late} $\mathtt{G_2}$ events cannot be always simply discarded like the case in simulating a deterministic consensus \citep{Daliot2006Agreement}.
Instead, each semi-synchronous consensus $\mathcal{B}_p'$ for each \emph{General} $p\in N$ is executed in an on-demand, silent (see \cite{Lamport1984Using} or later \cite{Lenzen2019AlmostasEasyas,Silence2018}) and preemptive way in responding to the $\mathtt{G_2}$ events of the self-stabilized $\mathcal P_{IA}$ primitives, as is shown in Fig.~\ref{algo:hop-think} and Fig.~\ref{algo:hop-simu}.
Here, to avoid any possible overlapping of the adjacent rounds in simulating the $\mathcal{B}$ protocol, the local time parameter $\delta_\mathtt{B}$ can be relaxed as $2\vartheta (\rho\Delta_\mathtt{B}+\epsilon_\mathtt{G}+d)$.
And the first $\delta_\mathtt{B}$ ticks waited in the $\mathcal{B}_p'$ function can be used to avoid any massages from formerly preempted $\mathcal{B}_p'$ being received in the currently initiated one.

\subsection{Basic Analysis}
For simplicity, in discussing all basic solutions, we set $\epsilon_\mathtt{G}=3d$, $\epsilon_\mathtt{B}=\rho\Delta_\mathtt{B}+\epsilon_\mathtt{G}+d$, $\delta_\mathtt{B}=2\vartheta \epsilon_\mathtt{B}$, $\Delta_\mathtt{B}=(K_\mathtt{B}+1)\delta_\mathtt{B}$, $\Delta_\mathtt{G}=\Delta_\mathtt{B}+4\vartheta d$, $\epsilon_\mathtt{H}=\epsilon_\mathtt{B}+\delta_\mathtt{B}$, and $\delta_3=\vartheta (\epsilon_\mathtt{H}+d)$.
It is easy to see that these equations can be trivially solved when $0\leqslant \rho\ll 1$.
Now as the properties of the underlying $\mathcal{B}$ protocol (required in (\ref{eq_B_Correctness1}), (\ref{eq_B_Correctness2}), and (\ref{eq_B_Agreement})) can be supported by at least the classical DBA protocols with $\eta_0=1$, here we construct the basic $\mathtt{Hoppelpopp}$ with these deterministic properties.
As $\eta_0=1$ in the basic solutions, we can remove the prefix \emph{Probabilistic-} from the name of every desired property of $\mathtt{Hoppelpopp}$.

\begin{lemma}
\label{lemma_live}
$\mathtt{Hoppelpopp}$ satisfies the \emph{Liveness} property required in (\ref{eq_promise_call1}) with $\Delta_0= \max\{2\varepsilon_0+\vartheta\Delta_\mathtt{A},\Delta_\mathtt{relax}\}+\Delta_\mathtt{v}+\Delta_\mathtt{B}+7d$.
\end{lemma}
\begin{IEEEproof}
Assume $e(\mathtt{H}\cup\mathtt{L_2},[t,t'],Q)=\emptyset$ with $t'\geqslant t+\Delta_0$.
Then by definition there is no happy bunny during $[t+2\varepsilon_0+\vartheta\Delta_\mathtt{A}+d,t']$.
So $e(\mathtt{G_0},I_2,Q)\in\mathbf{E_{gg}}$ holds for $I_2=t+\max\{2\varepsilon_0+\vartheta\Delta_\mathtt{A},\Delta_\mathtt{relax}\}+d+[0,\Delta_\mathtt{v}+d]$.
Denoting $t_p=Times(\mathtt{G_0},I_2,p)_{-1}$ for each $p\in Q$, as the condition in line~\ref{code_line_hop_init_do} of Fig.~\ref{algo:hop-init} would be true in all $q\in Q$ during $I_2+[0,d)$, we have $e(\mathtt{G_1},t_p+[0,d),Q)\in\mathbf{E_{gg}}$ in responding to $e(\mathtt{G_0},t_p,p)$.
So with the \emph{Extended Correctness 2} property of $\mathcal{P}_{IA}$, we have $e(\mathtt{G_2},I_3,Q)\in\mathbf{E_{gg}}$ with $I_3=t_p+[0,4d)$ in responding to $e(\mathtt{G_0},t_p,p)$.
Also, with the \emph{Extended Correctness 1} property of $\mathcal{P}_{IA}$, each $q\in Q$ would generate $\mathtt{G_3}$ with $c_p^{(q)}(t_q')=1$ at some $t_q'\in I_3$.
So with $\delta_\mathtt{B}\geqslant 2\vartheta (\rho\Delta_\mathtt{B}+\epsilon_\mathtt{G}+d)$ there is a simulation of $\mathcal{B}_p$ with the same input $1$ in all $q\in Q$.
So with the \emph{Correctness} property of the $\mathcal{B}$ protocol, this $\mathcal{B}_p$ would return $c_p^{(q)}=1$ in all $q\in Q$ and thus we have $e(\mathtt{H},I_4,Q)\in\mathbf{E_{gg}}$ with $I_4=I_3+\Delta_\mathtt{B}[\vartheta^{-1},1]+[0,d)\subseteq [t,t']$.
A contradiction.
\end{IEEEproof}

\begin{lemma}
\label{lemma_peaceability}
$\mathtt{Hoppelpopp}$ satisfies the \emph{Peaceability} property required in (\ref{eq_promise_call0}) with $\Delta_1=\Delta_\mathtt{v}+\Delta_\mathtt{B}+2d$.
\end{lemma}
\begin{IEEEproof}
As $\psi_q(t)\in \Psi_\mathtt{happy}$ holds for all $t_1\in [t,t']$, we have $e(\mathtt{G_0}\cup\mathtt{G_1},[t+d,t'],Q)=\emptyset$.
So with the \emph{Extended Unforgeability} property of $\mathcal{P}_{IA}$, we have $e(\mathtt{G_0}\cup\mathtt{G_1}\cup\mathtt{G_2},[t+d+\Delta_\mathtt{v},t'],Q)=\emptyset$ and thus $e(\mathtt{G},[t+d+\Delta_\mathtt{v}+\Delta_\mathtt{B}+d,t'],Q)=\emptyset$.
\end{IEEEproof}

\begin{lemma}
\label{lemma_probab}
$\mathtt{Hoppelpopp}$ satisfies the \emph{GoodLuck} property required in (\ref{eq_promise_come1}).
\end{lemma}
\begin{IEEEproof}
For every $t\in Times(\mathtt{H},[\Delta_\mathtt{c},+\infty),Q)$, by definition there is $e(\mathtt{H},t,q)\neq\emptyset$ with some $q\in Q$.
So the simulated $\mathcal{B}$ must have returned $c_p=1$ at some $t'\in[t-d,t]$ in $q$ and this simulated $\mathcal{B}$ is called in $q$ with satisfying the condition in line~\ref{code_line_hop_con} of Fig.~\ref{algo:hop-think}.
With the \emph{Extended $\Delta_\mathtt{agr}$-\emph{Relay}} property of $\mathcal{P}_{IA}$, $e(\mathtt{G_2},I',Q)\in\mathbf{E_{gg}}$ holds for some $I'$ with $t'\in I'$ and $|I'|\leqslant 2d$.
As all $q'\in Q$ not satisfying $\tau_{q'}\ominus\tau_{q'}^{p}\leqslant 4\vartheta d$ would set $c_p^{(q')}=0$ explicitly or silently during $I'+[0,d)$, with $c_p^{(q)}(t)=1$ there is at least one $q_1\in Q$ satisfying $\tau_{q_1}\ominus\tau_{q_1}^{p}\leqslant 4\vartheta d$ in calling the corresponding $\mathcal{B}_{p}'(1)$.
And with the \emph{Extended Uniqueness} property of $\mathcal{P}_{IA}$, the executions of $\mathcal{B}_{p}'$ would not be self-preempted since $\Delta_\mathtt{c}$ and the $\mathcal{B}$ protocol is simulated in $Q$.
So with the deterministic \emph{Agreement} property of $\mathcal{B}$, we have $Prob(\forall q\in Q:\exists t_q:|t-t_q|\leqslant \epsilon_\mathtt{H} \land e(\mathtt{H},t_q,q)\neq \emptyset )=1$.
\end{IEEEproof}

\begin{lemma}
\label{lemma_separate}
$\mathtt{Hoppelpopp}$ satisfies the \emph{Separation} property required in (\ref{eq_promise_come0}) with $\Delta_2=\epsilon_\mathtt{H}+\Delta_\mathtt{v}+\Delta_\mathtt{B}+d$ and $\Delta_\mathtt{relax}\geqslant \vartheta (\Delta_2+\Delta_3+\epsilon_\mathtt{H}+d)$.
\end{lemma}
\begin{IEEEproof}
For any $I\in \mathbf{I_\mathtt{H}}$, by definition $e(\mathtt{H},I,Q)\in\mathbf{E_{gg}}$ and $|I|\leqslant \epsilon_\mathtt{H}$ hold.
So by scheduling $\tau_\mathtt{relax}$ with $\Delta_\mathtt{relax}$, we have $e(\mathtt{G_1},I+[\epsilon_\mathtt{H}+d,\vartheta^{-1}\Delta_\mathtt{relax}-\epsilon_\mathtt{H}],Q)=\emptyset$.
So with the \emph{Extended Unforgeability} property of $\mathcal P_{IA}$, we have $e(\mathtt{G_2},I+[\epsilon_\mathtt{H}+d+\Delta_\mathtt{v},\vartheta^{-1}\Delta_\mathtt{relax}-\epsilon_\mathtt{H}],Q)=\emptyset$.
As $\Delta_\mathtt{relax}\geqslant \vartheta (2\epsilon_\mathtt{H}+\Delta_\mathtt{v}+\Delta_\mathtt{B}+\Delta_3+2d)$, we have $e(\mathtt{G_2},I+d+[\epsilon_\mathtt{H}+\Delta_\mathtt{v},\epsilon_\mathtt{H}+\Delta_\mathtt{v}+\Delta_\mathtt{B}+\Delta_3],Q)=\emptyset$ and thus $e(\mathtt{H},[t',t'+\Delta_3],Q)=\emptyset$ holds for $t'\in I+\epsilon_\mathtt{H}+\Delta_\mathtt{v}+\Delta_\mathtt{B}+d$.
\end{IEEEproof}

\subsection{Impossibility of Easier $\mathcal{B}$ Protocols}
Now with the deterministic properties of $\mathtt{Hoppelpopp}$, an optimistic idea might be that by relaxing the deterministic properties of the $\mathcal{B}$ protocols with allowing a probability $\eta_0\in(0,1]$, some easier probabilistic $\mathcal{B}$ protocol (with expected less rounds, message, or computation) might exist in satisfying (\ref{eq_B_Correctness1}), (\ref{eq_B_Correctness2}), and (\ref{eq_B_Agreement}) with some $\eta_0\in(0,1)$.
A more optimistic idea might even want to simplify the underlying $\mathcal{B}$ protocol into some fixed-round randomized protocol (such as the $Q_P$ protocol \citep{FM1989}, the $P_r$ protocol \citep{FM1997}, or some fixed-round randomized consensus routines \citep{Lenzen2019AlmostasEasyas}) that satisfies the agreement property with some positive probabilities for reaching expected-constant-time (ECT) P-SSBPS.
Here we show that all these apparent intuitions are over-optimistic under the discussed P-SSBPS framework.

Firstly, if the $\mathcal{B}$ protocol is replaced by some fixed-round randomized consensus routine $\mathcal{R}$ mentioned above, it is naive to think that the adversary would always expose itself to the unnecessary risk of reaching an agreement of $1$ for conducting a possible failure of the agreement in the execution of an $\mathcal{R}$ instance.
Namely, once we allow the adversary to acquire the ability to make some nonfaulty nodes deterministically output 0 while the other nonfaulty nodes output their values probabilistically (with the randomized common-coins), then the adversary can always repeat this semi-deterministic attack in every $\mathcal{R}$ instance without any cost, since a successful agreement of $0$ takes no further effect in the $\mathtt{Hoppelpopp}$ algorithm. However, the risks of the unsuccessful agreements would now be accumulated. In this situation, the adversary can take the strategy of initiating as many as possible such $\mathcal{R}$ instances without worrying about the outcome of any one of these instances being an agreement of 1 (as there is already at least one nonfaulty node outputting 0 in each such instance). And where the nonzero probabilities of unsuccessful agreements can be accumulated without limit, the ECT-SSBPS fails.
So, as such fixed-round randomized protocols cannot satisfy the \emph{GoodLuck} property required in (\ref{eq_promise_come1}), they cannot employed in the discussed P-SSBPS framework.

We can see that, by reducing the probability of unsuccessful agreement to be always $0$ (with an $(f+1)$-round DBA), all possible accumulations of such probabilities can be nullified when the underlying protocol is stabilized.
Another possible idea is to provide the $\mathcal{B}$ protocol with some $\eta_0\in(0,1)$.
At the first glance, by relaxing the deterministic properties to the probabilistic ones, the problem seems to be easier than DBA.
But here we show that this intuition is still false.

\begin{lemma}
\label{lemma_impossibility}
The problem of constructing $\mathcal{B}$ is as hard as the DBA problem.
\end{lemma}
\begin{IEEEproof}[sketch]
Suppose that $R$ is an easier $\mathcal{B}$ protocol with some $\eta_0\in(0,1)$.
As is required in the \emph{Deterministic Correctness 0} property of $\mathcal{B}$, no matter how $R$ is constructed, there must be some observable condition X0 for every nonfaulty node $i$ in executing $R$ such that when X0 is observed in the nonfaulty node $i$, $i$ would deterministically output $0$.
Besides, there exists the situation that X0 can be deterministically observed in every nonfaulty node.
And as is required in the \emph{Relaxed Correctness 1} property of $\mathcal{B}$, there exists the situation that every nonfaulty node does not observe X0 and outputs $1$ with a positive probability.

From these two properties we can see that, as there is the situation that X0 can be deterministically observed in each nonfaulty node, the observation of X0 cannot be probabilistic.
So, the adversary can deterministically make X0 being observed in only a part of the nonfaulty nodes.
For otherwise, the nonfaulty nodes would reach DBA on the observation of X0.
Now as is required in the \emph{Relaxed Agreement 1} property of $\mathcal{B}$, if there is a nonfaulty node that observes X0, there should be no nonfaulty node that would output $1$ with a positive probability in the same execution of $R$.
For otherwise, the \emph{Relaxed Agreement 1} property cannot be satisfied, as the adversary can deterministically make X0 being observed in only a part of the nonfaulty nodes in an infinite number of executions during which when a nonfaulty node outputs $1$ there is a $0$ probability that all nonfaulty nodes output $1$.

So, once a nonfaulty node observes X0, all nonfaulty nodes should deterministically output $0$. In this situation, as each node deterministically outputs $0$, the observation for this output cannot be probabilistic in each nonfaulty node. This means that each nonfaulty node deterministically observes X0. Thus, if one nonfaulty node observes X0, then all nonfaulty nodes observe X0. As there is the situation that each node does not observes X0 (required in \emph{Relaxed Correctness 1} property), the nonfaulty nodes can reach DBA on the observation of X0. A contradiction.
\end{IEEEproof}

It should be noted that in \cite{Lenzen2019AlmostasEasyas}, some fixed-round (or saying \emph{cut-off}) randomized synchronous consensus routine $C'$ \citep{FM1997} with the \emph{probabilistic agreement} property can be employed in constructing expected-sublinear-time SSBPS.
It is mainly because that a successful agreement of $0$ with the randomized $C'$ routine is also utilized in the SSBPS framework of \cite{Lenzen2019AlmostasEasyas} to indicate some abnormal state of the system, with which all nonfaulty nodes would be brought into a specific $\mathtt{RECOVER}$ state.
However, to ensure that the nonfaulty nodes in the $\mathtt{RECOVER}$ state would be synchronized (and thus the underlying synchronous routine $C'$ can be well simulated afterward) with the desired probability, the PS algorithm relies on the desired resynchronization points.
And these resynchronization points can only be supported by the recursively constructed resynchronization algorithms.
But in recursively constructing the resynchronization algorithms, the consensus routine $C'$ is still involved in generating the stabilized pulses for upper-layer PS.
At present, we do not know if and how the fixed-round randomized synchronous consensus routine like $C'$ can be employed in constructing peaceable expected-sublinear-time RSSBPS.

\section{Basic Solutions}
\label{sec:result}
\subsection{The Simple Solution}
As the $\mathcal{B}$ protocol is no easier than DBA, for the basic P-SSBPS solutions here we first deal with the case of $\mathcal{B}$ being some $(f+1)$-round immediate DBA.
For the simple solution, we assume the nonfaulty $\mathcal{S}$ runs the $\mathtt{Bunny}$ and the $\mathtt{Hoppelpopp}$ algorithms provided in Section~\ref{subsec:basic-Strategy}, \ref{subsec:realize-Absorption} and \ref{subsec:realize-Hoppelpopp} and set $\Delta_\mathtt{c}=\Delta_\mathtt{stb}+\Delta_\mathtt{relax}+\Delta_\mathtt{A}$, $\Delta_\mathtt{e}=\Delta_\mathtt{A}+\Delta_0+\Delta_1+\Delta_2+\Delta_3+\epsilon_2+\epsilon_0+d$, $\Delta_3= \Delta_\mathtt{A}$ and the other parameters according to the previous Lemmata.
Specifically, in the simple solution, we can reduce the parameter $\epsilon_\mathtt{H}$ to $\epsilon_\mathtt{B}$, since the agreement can always be reached at the $(f+1)$st round in immediate DBA.

\begin{theorem}
\label{theorem_result}
$\forall E^\chi\in \mathbf{E}:\exists t\leqslant \Delta_\mathtt{c}+\Delta_\mathtt{e}:E^\chi_Q[[t,+\infty)]\in \mathbf{E_1}$
\end{theorem}
\begin{IEEEproof}
For every $E^\chi\in \mathbf{E}$, with Lemma~\ref{lemma_live}, $e(\mathtt{H}\cup\mathtt{L_2},[t_0,t_0+\Delta_0],Q)\neq\emptyset$ holds for every $t_0\geqslant\Delta_\mathtt{c}$.
Firstly, if $e(\mathtt{H},[t_0,t_1],Q)=\emptyset \land e(\mathtt{L_2},t,Q)\neq\emptyset$ holds for some $t\in t_0+\epsilon_2+[0,\Delta_0]$ and $t_1= t+\Delta_\mathtt{A}+\epsilon_0+d+\Delta_1$, with Lemma~\ref{lemma_absorption_process}, $e(I',Q)\in\mathbf{E}_{\epsilon_0,T^-,T^+}^{|Q|,1+(l-1) K_\mathtt{A}}$, $\bar s_{\mathtt{L_2}}(e(\mathtt{L_2},I',Q))\in \mathbf{E}_{\epsilon_0+d,\phi_0^-,\phi_0^+}^{|Q|,l}$ and $(\mathtt{L_2},I',Q)_1\leqslant t+\Delta_\mathtt{A}$ hold for some $l\geqslant 1$, $I'=[t',t_1']$ with $t'\in [t,t+\Delta_\mathtt{A}]$ and $t_1'\in [t_1-\epsilon_0-d,t_1]$.
In this case, there exists $I_1\subseteq [t,t+\Delta_\mathtt{A}+\epsilon_0+d]$ satisfying $|I_1|\leqslant \epsilon_0+d\land e(\mathtt{L_2},I_1,Q)\in\mathbf{E_{gg}}$ and thus by definition of $\Psi_\mathtt{happy}$ we have $\forall q\in Q, t\in I_1+\epsilon_0+d+[0,\Delta_\mathtt{A}]: \psi_q(t)\in \Psi_\mathtt{happy}$.
So with Lemma~\ref{lemma_peaceability}, $e(\mathtt{G},[t_0'+\Delta_1,t_1],Q)=\emptyset$ holds with $t_0'=t+\Delta_\mathtt{A}+\epsilon_0+d$.
Denoting $t_2=Times(\mathtt{L_2},I',Q)_{-1}$ as the last $\mathtt{L_2}$ instant of $Q$ during $I'$, we have $\forall q\in Q, t\in [I_1+\epsilon_0+d,t_2+\Delta_\mathtt{A}+\epsilon_0+d]: \psi_q(t)\in \Psi_\mathtt{happy}$ and thus $e(\mathtt{H},[t_0,t_2+\Delta_\mathtt{A}+\epsilon_0+d],Q)=\emptyset$.
Again with Lemma~\ref{lemma_absorption_process}, $e(\mathtt{L_2},t_2',Q)\neq\emptyset$ also holds for some $t_2'\in[t_2,t_2+\Delta_\mathtt{A}]$.
So by iteratively applying Lemma~\ref{lemma_absorption_process}, we have $e(\mathtt{H},[t_0,+\infty),Q)=\emptyset$ and thus $e([t_3,+\infty),Q)\in \mathbf{E_1}$ for some $t_3\leqslant t+\Delta_\mathtt{A}+\Delta_1+\epsilon_0+d\leqslant t_0+\Delta_\mathtt{e}$.

Otherwise, if $e(\mathtt{H},[t_0,t_1],Q)=\emptyset \land e(\mathtt{L_2},[t_0+\epsilon_2,t_0+\epsilon_2+\Delta_0],Q)\neq\emptyset$ does not hold, we have $e(\mathtt{H},[t_0,t_4],Q)\neq\emptyset$ for $t_4=t_0+\epsilon_2+\Delta_0+\Delta_\mathtt{A}+\epsilon_0+d+\Delta_1$.
In this case, denoting $t_5$ as any an element in $(\mathtt{H},[t_0,t_4],Q)$ and $x_1=(\exists I\in \mathbf{I_\mathtt{H}}:t_5\in I)$, with Lemma~\ref{lemma_probab}, we have  $prob(x_1)\geqslant\eta_0=1$.
Assuming that $x_1$ is true for $I_2\in \mathbf{I_\mathtt{H}}$, with Lemma~\ref{lemma_separate}, there is $t_6\in I_2+\Delta_2$ satisfying $e(\mathtt{H},[t_6,t_6+\Delta_3],Q)=\emptyset$.
Denoting $t_7=Times(\mathtt{H},[t_5,t_6),Q)_{-1}$ and $x_2=(\exists I\in \mathbf{I_\mathtt{H}}:t_7\in I)$, with Lemma~\ref{lemma_probab}, we also have  $prob(x_2)\geqslant\eta_0=1$.
Now with the highest preemptive priority of the appearance of \emph{Hoppelpopp}, the premise of Lemma~\ref{lemma_initiation} holds.
And by iteratively applying Lemma~\ref{lemma_absorption_process}, we have $e([t_8,+\infty),Q)\in \mathbf{E_1}$ with $t_8\leqslant t_0+\Delta_\mathtt{e}$.
Thus, $Prob(\exists t'\leqslant t_8:e([t',+\infty),Q)\in \mathbf{E_1})\geqslant \eta_0=1$ holds.

So, in all cases we have $Prob(\exists t\in I:e([t,+\infty),Q)\in \mathbf{E_1})=1$ for all $I\subseteq [\Delta_\mathtt{c},+\infty)$ with $|I|\geqslant \Delta_\mathtt{e}$ and thus the conclusion holds.
\end{IEEEproof}
%

\subsection{ Early-Synchronizing Solution}
With the above proof, the basic solutions with early-stopping \citep{Toueg1987Fast,Dolev1990Early,Berman1992early} (or early-deciding \citep{Dolev2013EarlyDeciding}) DBA can be provided similarly (as the allowed errors for the coordinated $\mathtt{H}$ events are relaxed by $\epsilon_\mathtt{H}=\epsilon_\mathtt{B}+\delta_\mathtt{B}$).
The early-stopping basic solutions can be viewed as deterministic stabilization time optimization for the non-worst executions where the actual number of the Byzantine nodes is less than $f$.

\section{The Hybrid Solutions}
\label{sec:result2}
\subsection{A Hybrid Probabilistic Perspective}
Despite the impossibility shown in Lemma~\ref{lemma_impossibility}, we see that not all $\mathcal{B}$ protocols exclude fixed-round probabilistic routines.
For example, by integrating some expected-constant-round eventual RBA with some DBA routines, such as in \cite{Goldreich1990bestbothworlds} where the $K_0$-round synchronous RBA with $K_0=O(\log f)$ iterations is followed by an $(f+1)$-round synchronous DBA, the worst-case termination is guaranteed in $f+O(\log f)$ rounds while the expected number of rounds for termination is still constant (being independent of $f$).
Thus, by integrating some fixed-round randomized consensus routines (with some \emph{probabilistic agreement} property \citep{Lenzen2019AlmostasEasyas}) and some eventual early-stopping DBA protocols to construct the underlying $\mathcal{B}$ protocol, we can always ensure deterministic agreement with some $K_B= f+O(\log f)$ while to expect the agreement to be reached in constant time across at most two adjacent rounds.
Although this would introduce a slightly larger time error in the appearance of the \emph{Hoppelpopp}, it is still tractable with the two-stage absorption process.
With this, the probabilistic optimization for the P-SSBPS problem becomes quite interesting.

\subsection{The Optimistic Side}
Now, to explore the possible optimizations for the stabilization time, the following discussion can only be presented in a relaxed informal way.
The formal analysis of all these possibilities cannot be covered in this paper.
By this adventure, we hope that, despite the rigorous analysis, the readers would make a better understanding of the underlying problems, limitations, and trade-offs behind the provided basic ideas and solutions.

Firstly, in the most common sense, providing that the underlying protocols and the timers employed in the $\mathtt{Hoppelpopp}$ algorithm are all stabilized since $\Delta_{\mathtt{c}}$, the expected extra time $\Delta_{\mathtt{e}}$ for reaching P-SSBPS can be constant (being independent of $f$) in the presence of $f$ actual Byzantine nodes.
This is because that the actual execution time of a $\mathcal{B}$ instance is expected to be constant even in the presence of the maximal number of actual Byzantine nodes.
Notice that it is orthogonal to the deterministic non-worst-case time optimization of early-stopping DBA where the number of actual Byzantine nodes is less than $f$.

For the unstabilized duration before $\Delta_{\mathtt{c}}$, firstly, from the most optimistic perspective, if the \emph{I-Accept} primitives and the timers $\tau_\mathtt{v}$ and $\tau_\mathtt{relax}$ can be stabilized in ECT, it is easy to see that the system would be stabilized in ECT if there is no unstabilized $\mathcal{B}$ instance outputting $1$.
So in this situation, the adversary can only postpone the ECT stabilization by generating outputs of unstabilized $\mathcal{B}$ instances.
Assume that there are at most $f+1$ such unstabilized $\mathcal{B}$ instances (this can be easily done with explicitly allowing only $f+1$ nodes being the possible \emph{Generals}).
In this case, for the possible unsynchronized (the simulated rounds might be not time-aligned) unstabilized $\mathcal{B}$ instances, we can restrict the power of the adversary by locally checking and maintaining the ongoing states of the $\mathcal{B}$ instances.
For example, the round-messages of each ongoing $\mathcal{B}$ instance can be locally checked in that when the $k$th-round-messages are received from less than $n-f$ distinct nodes during the $k$th round of the $\mathcal{B}$ instance in a nonfaulty node $i$, $i$ can abnormally terminate this $\mathcal{B}$ instance with the default output $0$.
Meanwhile, we can employ some FTA functions in each ongoing $\mathcal{B}$ instance to bring the beginning instants of the rounds of at least $n-2f$ nonfaulty nodes into unison in the progress of this $\mathcal{B}$ instance.
So, in the worst case, the execution of this $\mathcal{B}$ instance can be maintained in some $n-2f$ nonfaulty nodes until the last round of the partially synchronized instance.
And further strategies can even bring the remaining $f$ nonfaulty nodes to the partially synchronized instance and then make it to be a wholly synchronized one.
In this situation, if we can optimistically assume that these instances (at most $f+1$ ones) terminate in random instants, i.e., with equal possibilities during $[0,\Delta_{\mathtt{c}}]$, then there is a probability $((\Delta_{\mathtt{c}}-\Delta_{\mathtt{d}})/\Delta_{\mathtt{c}})^{(f+1)}$ that no unstabilized $\mathcal{B}$ instance outputs during $[t,t+\Delta_{\mathtt{d}}]$ for any $t\in [0,\Delta_{\mathtt{c}}]$.
In this case, there is a constant probability (towards $e^{-x}$ when $f\to +\infty$) that no unstabilized $\mathcal{B}$ instance outputs during any an $O(x\Delta_{\mathtt{c}}/f)$ duration in $[0,\Delta_{\mathtt{c}}]$.
With this, some kind of \emph{pseudo-synchronization} pattern can be provided to allow some short-term critical tasks to be performed before the overall stabilization of the system.

Otherwise, if the termination instants of the unstabilized $\mathcal{B}$ instances cannot be assumed to be randomly distributed, by allowing $n\gg f$, we can take some early-stoping version of the phase-king protocol \citep{Berman1992early} (with the overwhelming majority configured as $n-2f$ or less) with each possible \emph{General} being assigned with disjoint phase-kings.
As there are only $f+1$ possible \emph{Generals}, this can be done with $n=\Omega (f^2)$.
With this, in keeping an ongoing $\mathcal{B}$ instance from being early-stopped before $k$ rounds of time, the adversary should at least invest $k$ faulty nodes as the last $k$ phase-kings in this $\mathcal{B}$ instance.
And if the adversary invests no faulty node in an unstabilized $\mathcal{B}$ instance, this instance would be early-stopped in constant time (as there would be a nonfaulty phase-king in each such round) with simulating the early-stoping phase-king protocol.
In this situation, to prevent the formation of the \emph{pseudo-synchronization} pattern before $kO(\delta_{\mathtt{B}})$, the adversary should at least invest $k+(k-1)+\cdots+2+1=k(k+1)/2$ faulty nodes.
As there are at most $f$ faulty nodes, the adversary can at most contaminate the first $O(\sqrt{f}\delta_{\mathtt{B}})$ time or there is a constant probability that the \emph{pseudo-synchronization} pattern forms before $O(\sqrt{f}\delta_{\mathtt{B}})$ in the unstabilized system.
And once the adversary chooses to contaminate the first $O(\sqrt{f}\delta_{\mathtt{B}})$ time, the system can be stabilized in $O(\sqrt{f})$ time as $\delta_{\mathtt{B}}$ is constant.
But notice that this is under the assumption of the ECT stabilization of the \emph{I-Accept} primitives and the timers $\tau_\mathtt{v}$ and $\tau_\mathtt{relax}$.
Obviously, this assumption is over-optimistic in considering the malicious adversary.

In considering the full-capacity adversary, firstly, for the ECT stabilization of the timer $\tau_\mathtt{v}$, as the adversary can always control the situation until the last round of the $\mathcal{B}$ instance with some positive probability, a linear time for executing each $\mathcal{B}$ instance is needed to be scheduled.
In this situation, the full version of the self-stabilizing BA solution provided in \cite{Daliot2006Agreement} can be employed where the executions of the BA are initiated by indexed \emph{I-Accept} primitives.
Concretely, each initiation of the BA protocol with the same \emph{General} can be differentiated with a distinct index $m_0$ and make it as the parameter $m$ of the \emph{I-Accept} primitive.
For the simplest example, by setting $m_0=(m_0+1)\bmod M_0$ in each initiation of the indexed \emph{I-Accept} primitive (with a finite but sufficiently large $M_0$), the scheduled timeout of $\tau_\mathtt{v}$ in the $\mathtt{Hoppelpopp}$ algorithm can be set as a constant number ($\Delta_0=13d$ in \cite{Daliot2006Agreement}).
Although this is at the expense of more parallel executions of the $\mathcal{B}$ protocol (an extra $O(f)$ factor in message and computation complexity), it is in line with our main direction (i.e., to trade abundant bandwidth for stabilization time).

And secondly, for the ECT stabilization of the \emph{I-Accept} primitive, we would have to take a closer investigation into the indexed \emph{I-Accept} primitives.
Concretely, by setting an ascending index $m_0$ in the initiation of each new indexed \emph{I-Accept} primitive, at any time there should be at most $K_\mathtt{c}=O(f)$ parallel $\mathcal{B}$ instances running for the same \emph{General} with ascending indexes.
So when there are more than $K_\mathtt{c}$ ongoing $\mathcal{B}$ instances or the indexes of them are not strictly $1$-by-$1$ ascending for the same \emph{General} $j$ in any nonfaulty node $i$, $i$ can clear all the on-going $\mathcal{B}$ instances together with all the on-going \emph{I-Accept} primitives in $i$, since it is an unstabilized state.
Thus, at instant $0$, we can assume that there are at most $K_\mathtt{c}$ ascending-indexed on-going $\mathcal{B}$ instances running for each \emph{General} $j$ in every nonfaulty node $i$, with each on-going $\mathcal{B}$ instance being assigned with disjoint phase-kings (with $n=\Omega(K_\mathtt{c} f^2)=\Omega(f^3)$).

Now, as an on-going $\mathcal{B}$ instance can block new \emph{I-Accept} primitive with the same parameter $m$ (see \cite{Daliot2006Agreement} for details), when a nonfaulty node $j$ initiates a new \emph{I-Accept} primitive (with $j$ being the \emph{General}), besides taking $m_0=(m_0+1)\bmod M_0$ as the index of the new \emph{I-Accept} primitive, $j$ can also randomly choose a base-number $m_1$ from some finite set $[0:M_1-1]$ and then set $m=m_1 M_0+m_0$ as the parameter of the new \emph{I-Accept} primitive.
With this, there can be a high possibility that the adversary cannot block the new \emph{I-Accept} primitive initiated in nonfaulty nodes by setting the initial states of the nonfaulty nodes with $K_\mathtt{c}$ on-going $\mathcal{B}$ instances for each \emph{General}, providing that $M_1$ is sufficiently large.
Concretely, by setting $M_1=\Omega(n^2 K_\mathtt{c})$, there is a at least a probability $\Omega((1-1/n)^{n})$ that the randomly chosen base-number of the new \emph{I-Accept} primitive initiated in each nonfaulty \emph{General} is not in the base-numbers of the on-going $\mathcal{B}$ instances running in all nonfaulty nodes.
Meanwhile, the assigned phase-kings can remain unchanged with respect to $m_0$.
Also, in our use of the \emph{I-Accept} primitives, when the execution of an \emph{I-Accept} primitive does not terminate in $8d$, it can be immediately reset, since the acceptance of such an \emph{I-Accept} primitive would take no effect in the $\mathtt{Hoppelpopp}$ algorithm.
Thus, with a sufficiently large $M_1$ (by adding $\Omega(\log(n^2f))$ bits in each message), almost all \emph{I-Accept} primitives (excluding the ones initiating on-going $\mathcal{B}$ instances) can be stabilized in constant time.
And with a high probability that the new \emph{I-Accept} primitives would not be blocked by the ongoing ones.

For optimizing the timeout of the timer $\tau_\mathtt{relax}$, firstly, with the deterministic guarantee for the worst cases, instead of always scheduling the timeout $\Delta_\mathtt{relax}$ as a constant in the $\mathtt{Hoppelpopp}$ algorithm, we can dynamically set $\Delta_\mathtt{relax}$ according to the actual execution time of the terminated $\mathcal{B}$ instance.
Namely, $\Delta_\mathtt{relax}$  can be set as some $\Delta_\mathtt{G} + O(\tau_\mathtt{B})$ with $\tau_\mathtt{B}$ being the actual execution time of the terminated $\mathcal{B}$ instance.
So when all the underlying primitives are stabilized, an $\mathcal{B}$ instance terminated with output $1$ would guarantee that the currently buffered $\mathcal{B}$ instances are the only remaining $\mathcal{B}$ instances to be handled before the stabilization of the system.
However, when the underlying primitives are not stabilized, the adversary can still set $\Delta_\mathtt{relax}=\Omega(f)$.
For ECT stabilization of $\tau_\mathtt{relax}$, by scheduling $\Delta_\mathtt{relax}$ being always independent of $f$, for example taking some value $\Delta_\mathtt{G} +c\delta_\mathtt{B}$ with a constant $c<K_0$, once the $\mathcal{B}$ instance reaches agreement of $1$ within $c\delta_\mathtt{B}$ time from its initiation, then the system would be stabilized, if all the underlying primitives are stabilized.
But this is at the expense of overall randomized stabilization time without the deterministic guarantee.

\subsection{Obstacles of ECT-SSBPS}
It should be noted that, however, all the possible optimization for the stabilization time informally discussed above are in the context of some non-worst cases in the whole picture.
Yet we do not know if there is an ECT-SSBPS solution under the classical static and \emph{strong} adversary \citep{Dolev2014PulseGeneration}.
As is informally discussed, in considering ECT-SSBPS under such an adversary, we think that the main obstacle is the stabilization of the underlying $\mathcal{B}$ instances.
Namely, the system can be \emph{pseudo-synchronized} \emph{before} the stabilization of all on-going $\mathcal{B}$ instances.
And when the system is \emph{pseudo-synchronized}, it can still be intervened by the later terminations of the possible unstabilized on-going $\mathcal{B}$ instances configured by the adversary at instant $0$.
To this end, although there are some tricks (discussed all above) in restricting the power of the adversary from conducting undesired outputs with these unstabilized $\mathcal{B}$ instances, yet we can only expect a rather relaxed synchronization pattern in the unstabilized duration $[0,\Delta_{\mathtt{c}}]$ such that some \emph{pseudo-synchronization} would appear in some $\Omega(f-\sqrt{f})$ duration in expected $O(\sqrt{f})$ time with $n=\Omega(f^3)$.
This might be helpful in considering some real-world applications, such that some short-term critical tasks might be done with the so-called \emph{pseudo-synchronized} system state before the overall deterministic stabilization.
While the problem of the existence or nonexistence of an ECT-SSBPS solution, is not decided yet.

\subsection{Some Possible Composition}
The so-called \emph{pseudo-synchronization} in the unstabilized P-SSBPS system might help provide some kind of sublinear-time resynchronization points \citep{Lenzen2019AlmostasEasyas,Dolev2014PulseGeneration}.
For example, by employing some variants of the P-SSBPS solution provided here as an underlying resynchronization protocol, the resynchronization algorithm in \cite{Dolev2014PulseGeneration} might be replaced for faster peaceable RSSBPS (P-RSSBPS).
To this, the main problem is the integration of the provided resynchronization points with the deterministic layers of \emph{FATAL} \citep{Dolev2014PulseGeneration}.
As the provided P-RSSBPS and \emph{FATAL} are all peaceable, the compositional result is also peaceable.
Another example could be to integrate the resynchronization points with the highest-layer synchronization algorithm employed in \cite{Lenzen2019AlmostasEasyas}.
Namely, by replacing the recursively constructed highest-layer resynchronization algorithm, at least the original linear-time factor $2\cdot 4\cdot 5$ in \cite{Lenzen2019AlmostasEasyas} is avoided.
However, as the original SSBPS solutions provided in \cite{Lenzen2019AlmostasEasyas} are not peaceable, the compositional results are not peaceable.

For better compositions of the randomized stabilization time and the deterministic worst-case linear one, an unsolved problem is the scheduling of the timer $\tau_\mathtt{relax}$.
Namely, when the timeout $\Delta_\mathtt{relax}$ is set as $O(f)$, P-DSSBPS solutions can be derived with the deterministic $O(f)$ stabilization time.
But the adversary can leverage this to prevent the initiation of new $\mathcal{B}$ instances when the system is not stabilized.
However, if we only allow $\Delta_\mathtt{relax}$ being scheduled as $o(f)$, only the P-RSSBPS solution can be derived in considering the worst cases (when the execution time of all the $\mathcal{B}$ instances are greater than $\Delta_\mathtt{relax}$).
So, in considering the worst cases, the timer $\tau_\mathtt{relax}$ should always be scheduled with a $O(f)$ timeout.
But with this, the \emph{pseudo-synchronized} state can only be derived in some non-worst cases.

\section{Other Discussion}
\label{sec:discussion}
In the previous sections, we have presented a basic P-SSBPS solution based on the absorption processes and the emergency processes in the bounded-delay message-passing systems.
This basic solution assumes a non-cryptographic, Byzantine, and static adversary.
In some benign settings, some kinds of \emph{weaker} adversary can be assumed in handling the corresponding P-SSBPS problems.
While in the opposite direction, an \emph{adaptive} adversary \citep{King2011Breaking} can arbitrarily choose up-to $f$ nodes from $N$ being Byzantine nodes and change them over time.
In considering the various kinds of adversaries, here we first discuss how the basic P-SSBPS solution can be extended to these cases under the same framework.
Then, we shortly discuss the relation between P-SSBPS and peaceable digital clock synchronization.
Last we discuss some utilities of the provided P-SSBPS solutions and then make a comparison with existing SSBPS solutions.

\subsection{For the Weaker Adversaries}
Under the basic P-SSBPS framework, the decoupled absorption process and emergency process can be discussed separately.
For the absorption process, with the basic strategy presented in Section~\ref{sec:overview}, desired properties can be provided with sparse resource occupation under the omniscient static adversary.
So it is trivial in considering the weaker adversaries.
For the emergency process, firstly, if the faulty nodes are only allowed to generate consistent messages (with reliable broadcast), constant-time stabilization of P-DSSBPS is trivial, since it is trivial to reach a one-round agreement in the semi-synchronous simulation.
Secondly, by assuming a cryptographic adversary, we can use authenticated BA \citep{Dolev1983Authenticated,Gupta2010Authenticated,AbrahamDolev2019,Abraham2019optimal} to construct easier $\mathcal{B}$ protocols (basic or hybrid).
Also, by assuming that the faulty nodes can only be randomly chosen \citep{MAURER20143153}, the early-stopping phase-king protocols can trivially reach both deterministic $(f+1)$-round termination and expected-constant-round termination with $n=\Omega(f^3)$.
In this setting, as the unstabilized $\mathcal{B}$ instances can also terminate in ECT, ECT-SSBPS can be trivially achieved.

\subsection{For the Dense-Time Adaptive Adversary}
Besides, the provided P-SSBPS solutions can also work under the \emph{adaptive} adversary, providing that the transformations between the faulty and nonfaulty nodes are not over-overwhelming nor over-frequent in the dense time.
Concretely, when a faulty node $p$ transforms into a nonfaulty node at an instant $t_p$, $p$ cannot be immediately viewed as a node in $Q$.
Instead, $p$ would be viewed as a special kind of node (the transforming node) during $[t_p,t_p+\Delta_\mathtt{trans}]$, where the transforming duration $\Delta_\mathtt{trans}\geqslant\Delta_\mathtt{c}+2\Delta_\mathtt{e}$ should be sufficiently long.
And denoting $F(t)$ and $\tilde{Q}(t)$ as the set of all faulty and transforming nodes at $t$, respectively, $|F(t)\cup\tilde{Q}(t)|\leqslant f$ should be satisfied in a nonfaulty $\mathcal{S}$ for all $t\in[0,+\infty)$.
With this, if $\mathcal{S}$ is not stabilized at $t$, with $\Delta_\mathtt{trans}\geqslant\Delta_\mathtt{c}+2\Delta_\mathtt{e}$, $|Q\cup\tilde{Q}_\mathtt{c}|\geqslant n-f$ holds during every $2\Delta_\mathtt{e}$ time interval since $t$, where $\tilde{Q}_\mathtt{c}(t)\subseteq\tilde{Q}(t)$ is the set of the nodes which have been in $\tilde{Q}$ for at least $\Delta_\mathtt{c}$ time at $t$.
And if $\mathcal{S}$ is stabilized at $t$, each node in $\tilde{Q}(t)$ would be stabilized by the $n-f$ nonfaulty nodes before it can be included in $Q$.
So, by allowing $Q$ as $Q\cup\tilde{Q}_\mathtt{ce}$, where $\tilde{Q}_\mathtt{ce}(t)\subseteq\tilde{Q}_\mathtt{c}(t)$ is the set of the nodes which have been in $\tilde{Q}_\mathtt{c}$ for at least $\Delta_\mathtt{e}$ time at $t$, a similar result of Theorem~\ref{theorem_result} can still be derived under such a restricted dense-time adaptive adversary.
Actually, in dense-time systems, the ability of the \emph{adaptive} adversary in choosing different faulty nodes in any round-based systems should also be restricted by a bounded number (of at most $f$) and a bounded frequency (of at least a maximal round cycle).
Thus, there is no essential difference between the synchronous adaptive adversary \citep{King2011Breaking} and the restricted dense-time adaptive adversary discussed here in the context of the SSBPS problem.

\subsection{Peaceable Dense-Time Clock Synchronization}
By setting the range of the absorption counter $K_\mathtt{A}$ as a specific number, when the system is stabilized, the synchronized absorption cycles of the absorption processes can be directly used as synchronized digital clock cycles, with which peaceable digital clock synchronization (and thus peaceable dense-time clock synchronization) is also established in the P-SSBPS system.
Thus, peaceable (dense-time) clock synchronization can be achieved in the bounded-delay model without any extra overhead.
But instead of converting some digital clock synchronization algorithms (such as the expected-constant-round ones in \cite{Ben2008Fast}) to dense-time clock synchronization, here the SSBPS algorithm is directly used as the dense-time clock synchronization algorithm.
This provides an interesting supplement to the question posed in \cite{Ben2008Fast} whether digital clock synchronization algorithms can be conveniently converted to dense-time clock synchronization algorithms.

The peaceable dense-time clock synchronization with the two-stage absorption processes can be further optimized in many ways.
Firstly, digital clock synchronization with small clock cycles can be handled in P-SSBPS with the strategies provided in \cite{daliot2006self,Dolev2014PulseGeneration}.
And for digital clock synchronization with large clock cycles, for example, an extra integer can be added in the pulse message during the first stage of the absorption process to indicate the current value of the upper-layer clock.
With this, the stabilization time of the dense-time clock synchronization with large clock cycles can be independent of the upper-layer clock cycle.

\subsection{Performance and Utilities}
By trading the system resources reserved for the higher-layer applications for the desired stabilization time of SSBPS, the P-SSBPS solutions provided here have particular usefulness in systems with heavy synchronous communication, computation, etc, in the application layers.
Firstly, if the system is not stabilized, as long as the system bandwidth is not exhausted, the unused portion can be temporarily allocated for reaching SSBPS as fast as possible.
And once the system is stabilized, the resources occupation of the P-SSBPS would be lowered to a similar level of the approximate agreement with constant-bits messages.
Concretely, in the basic solutions, the two-stage absorption process needs only one extra bit in representing the two mark values ($\{\mathtt{GOOD,BEST}\}$ and $\{\mathtt{GOOD}\}$) in the stabilized system.
And the temporal trails can be logically observed for the desired duration with local timers instead of continuous observations.
With these, the formerly occupied extra system resources can now be released and allocated to the higher-layer applications in stabilized systems.
For example, each happy bunny $q$ can observe the $\mathtt{G_2}$ events in deciding if the system-wide stabilization is established.
If there is no $\mathtt{G_2}$ event being generated in $q$ for a sufficiently long time, $q$ knows that all bunnies are happy (and all be the best) and the temporarily allocated resources can all be released and the higher-layer synchronous protocols can be initiated and start to run.
In a bigger picture, the provided P-SSBPS framework is also compatible with network topology reconstruction \citep{MAURER20143153} and other system-recovery schemes.
For example, when the SSBPS system \emph{cannot} be stabilized under some ill-formed network topology, an agreement of $1$ in the emergency process can indicate the beginning of a topology reconstruction process (being included in the emergency process) during which the network topology can be reconstructed in favor of P-SSBPS.
And after the reconstruction of the network topology, if the SSBPS system is stabilized, no more topology reconstruction process would be performed, with which the topology reconstruction is also peaceable.

Comparing with state-of-the-art solutions, firstly, the DSSBPS solutions provided in \cite{Lenzen2019AlmostasEasyas} have the complexity of DBA even when the system is stabilized.
The peaceable DSSBPS solution provided here reduces the message and computation complexity in stabilized SSBPS to that of approximate agreement with $2$-bit round-messages, which is optimal in the presence of Byzantine faults.
And the linear stabilization time required in the peaceable DSSBPS is approximately $\Delta_\mathtt{stb}+5\Delta_\mathtt{B}$ in considering the linear coefficient and can be further improved.
Also, the worst-case execution time $\Delta_\mathtt{B}$ of BA is only $(f+1)\delta_\mathtt{B}$ where $\delta_\mathtt{B}$ is approximately $8d$ (relatively smaller than that of \cite{Lenzen2019AlmostasEasyas} and can be further improved).
Secondly, in considering RSSBPS solutions, although the solutions provided in \cite{Dolev2014PulseGeneration} also support peaceability, the stabilization time is expected-linear with a significant linear coefficient.
And although the RSSBPS solutions provided in \cite{Lenzen2019AlmostasEasyas} can achieve expected-sublinear-time stabilization, it does not support peaceability.
Here we argue that, in considering the overall efficiency of SSBPS in practical applications, as the distributed systems should also perform various upper-layer synchronous activities, the efficient DSSBPS solutions with sparse resource occupation in the stabilized states are even better than the ECT solutions without it.
Thirdly, in hybrid SSBPS solutions, the expected stabilization time can be optimized with the guarantee of deterministic linear-time stabilization.
These optimizations can be helpful not only in reaching faster stabilization with a significant probability in real-world applications but also in reaching expected-sublinear-time stabilization in compositional SSBPS (including P-SSBPS) solutions.

\section{Conclusion and Future Work}
\label{sec:conclusion}
In this paper, we have investigated the P-SSBPS problem and presented a general solution to the problem with the decoupled absorption process and emergency process.
In all provided P-SSBPS solutions, the two-stage absorption process is built upon the coarser preemptive coordination and the finer non-self-stabilizing approximate agreements, with observations of several temporal trails of the pulsing processes.
With this, in the stabilized P-SSBPS system, as there is no consensus being performed, the message complexity and computation overhead are all similar to the approximate agreement.
In the basic P-DSSBPS solutions, the emergency process is built upon the self-stabilizing DBA with or without the integration of early-stopping properties.
Analysis shows that deterministic linear-time stabilization can be achieved in the basic P-DSSBPS solution with optimized linear coefficients.
In the hybrid P-SSBPS solutions, some probabilistic time-optimizations are also informally discussed.
These optimizations can help reach expected-sublinear-time stabilization in compositional SSBPS (peaceable or not) solutions.
In the extended P-SSBPS solutions, the emergency processes can be specifically designed for various adversaries under the general framework.
And the effect of transformations between faulty and nonfaulty nodes is also briefly discussed with the restricted dense-time adaptive adversary.
Also, digital clock synchronization can be naturally achieved in the provided P-SSBPS solution without any extra expense.
And other system-recovery schemes such as the peaceable topology reconstruction are also supported in the basic P-SSBPS framework.
And in formalization, the notations and analysis are all under the formal representation of the temporal processes and temporal trails.
This gives a self-contained description of the dense-time SSBPS problem and would also facilitate future optimization of the P-SSBPS solutions.

Despite the advantages, a main disadvantage of the basic P-SSBPS solution is the complete-graph assumption of the communication network, which means the highest network connectivity and node degrees.
To break this, as the almost everywhere (a.e.) BA can be reached in some bounded-degree networks \citep{Dwork1986,Upfal1992}, we can see that at least the a.e. absorption process can also be reached by simulating the absorption process in such networks.
Further, inspired by \cite{King2009}, here it is interesting to ask if the absorption process can be efficiently reached in bounded-degree networks and how the absorption process (or at least the a.e. absorption process) can be deterministically maintained peaceably.
As some emerging HRT-SCS like the large-scale moving swarms are often with strict restrictions on the average energy consumption and the maximal communication ranges, the peaceability property of SSBPS in the bounded-degree networks is of practical significance.
And with the integration of network topology reconstruction under the P-SSBPS framework, we wonder if desired network topology can be efficiently reconstructed in favor of P-SSBPS.
Also, inspired by \cite{OptimalGradientDynamic2010,Gradient2019}, we wonder if the P-SSBPS solutions can be integrated with some specific synchronization requirements (such as the gradient clock synchronization \citep{Gradient2004}) or some new failure-restricted synchronization schemes (such as the mutex propagation strategy \citep{YuCOTS2021}) in real-world popular networks \citep{Leighton1992On}.
Lastly, the stabilization time of the P-SSBPS solutions provided in this paper is restricted by the stabilization of the underlying primitives.
We wonder if P-SSBPS can be reached with less stabilization time than that of the self-stabilizing primitives employed in this paper.

\bibliographystyle{IEEEtran}
\bibliography{IEEEabrv,PSSBPS}

\end{document}